\documentclass[journal, onecolumn,12pt]{IEEEtran}
\IEEEoverridecommandlockouts
\usepackage{cite}
\usepackage{amsmath,amssymb,amsfonts}
\usepackage{algorithmic,algorithm}
\usepackage{graphicx}
\usepackage{caption}
\usepackage{subcaption}
\usepackage[margin=1in]{geometry}
\usepackage{setspace} 

\usepackage{cleveref}
\usepackage{textcomp}
\usepackage{xcolor}
\usepackage{bbm}
\usepackage{tikz}
\usepackage{color,soul}
\usetikzlibrary{patterns}

\newtheorem{theorem}{Theorem}
\newtheorem{lemma}{Lemma}
\newtheorem{proposition}{Proposition}
\newtheorem{definition}{Definition}
\newtheorem{corollary}{Corollary}
\newtheorem{remark}{Remark}
\newtheorem{assumption}{Assumption}
\newtheorem{example}{Example}

\def\BibTeX{{\rm B\kern-.05em{\sc i\kern-.025em b}\kern-.08em
    T\kern-.1667em\lower.7ex\hbox{E}\kern-.125emX}}
\begin{document}

\title{Online Energy Minimization Under a Peak Age of Information Constraint                       
\thanks{We acknowledge support of the Department of Atomic Energy, Government of India, under project no. RTI4001.} \\
\thanks{This article was presented in part at the Proceedings of WiOpt 2021.}
}
\author{\IEEEauthorblockN{Kumar Saurav and Rahul Vaze}
	\IEEEauthorblockA{\\ \textit{School of Technology and Computer Science} \\
		\textit{Tata Institute of Fundamental Research}\\
		Mumbai, India. \\
		kumar.saurav@tifr.res.in, rahul.vaze@gmail.com} 
}

\vspace{-1in}
\def\onehalf{\frac{1}{2}}
\def\etal{et.\/ al.\/}
\newcommand{\bydef}{\triangleq}
\newcommand{\tr}{{\it{tr}}}
\def\SNR{{\textsf{SNR}}}
\def\bydef{:=}
\def\bba{{\mathbb{a}}}
\def\bbb{{\mathbb{b}}}
\def\bbc{{\mathbb{c}}}
\def\bbd{{\mathbb{d}}}
\def\bbee{{\mathbb{e}}}
\def\bbff{{\mathbb{f}}}
\def\bbg{{\mathbb{g}}}
\def\bbh{{\mathbb{h}}}
\def\bbi{{\mathbb{i}}}
\def\bbj{{\mathbb{j}}}
\def\bbk{{\mathbb{k}}}
\def\bbl{{\mathbb{l}}}
\def\bbm{{\mathbb{m}}}
\def\bbn{{\mathbb{n}}}
\def\bbo{{\mathbb{o}}}
\def\bbp{{\mathbb{p}}}
\def\bbq{{\mathbb{q}}}
\def\bbr{{\mathbb{r}}}
\def\bbs{{\mathbb{s}}}
\def\bbt{{\mathbb{t}}}
\def\bbu{{\mathbb{u}}}
\def\bbv{{\mathbb{v}}}
\def\bbw{{\mathbb{w}}}
\def\bbx{{\mathbb{x}}}
\def\bby{{\mathbb{y}}}
\def\bbz{{\mathbb{z}}}
\def\bb0{{\mathbb{0}}}

\def\bydef{:=}
\def\ba{{\mathbf{a}}}
\def\bb{{\mathbf{b}}}
\def\bc{{\mathbf{c}}}
\def\bd{{\mathbf{d}}}
\def\bee{{\mathbf{e}}}
\def\bff{{\mathbf{f}}}
\def\bg{{\mathbf{g}}}
\def\bh{{\mathbf{h}}}
\def\bi{{\mathbf{i}}}
\def\bj{{\mathbf{j}}}
\def\bk{{\mathbf{k}}}
\def\bl{{\mathbf{l}}}
\def\bm{{\mathbf{m}}}
\def\bn{{\mathbf{n}}}
\def\bo{{\mathbf{o}}}
\def\bp{{\mathbf{p}}}
\def\bq{{\mathbf{q}}}
\def\br{{\mathbf{r}}}
\def\bs{{\mathbf{s}}}
\def\bt{{\mathbf{t}}}
\def\bu{{\mathbf{u}}}
\def\bv{{\mathbf{v}}}
\def\bw{{\mathbf{w}}}
\def\bx{{\mathbf{x}}}
\def\by{{\mathbf{y}}}
\def\bz{{\mathbf{z}}}
\def\b0{{\mathbf{0}}}
\def\opt{\mathsf{OPT}}
\def\on{\mathsf{ON}}
\def\off{\mathsf{OFF}}
\def\bA{{\mathbf{A}}}
\def\bB{{\mathbf{B}}}
\def\bC{{\mathbf{C}}}
\def\bD{{\mathbf{D}}}
\def\bE{{\mathbf{E}}}
\def\bF{{\mathbf{F}}}
\def\bG{{\mathbf{G}}}
\def\bH{{\mathbf{H}}}
\def\bI{{\mathbf{I}}}
\def\bJ{{\mathbf{J}}}
\def\bK{{\mathbf{K}}}
\def\bL{{\mathbf{L}}}
\def\bM{{\mathbf{M}}}
\def\bN{{\mathbf{N}}}
\def\bO{{\mathbf{O}}}
\def\bP{{\mathbf{P}}}
\def\bQ{{\mathbf{Q}}}
\def\bR{{\mathbf{R}}}
\def\bS{{\mathbf{S}}}
\def\bT{{\mathbf{T}}}
\def\bU{{\mathbf{U}}}
\def\bV{{\mathbf{V}}}
\def\bW{{\mathbf{W}}}
\def\bX{{\mathbf{X}}}
\def\bY{{\mathbf{Y}}}
\def\bZ{{\mathbf{Z}}}
\def\b1{{\mathbf{1}}}

\def\bbA{{\mathbb{A}}}
\def\bbB{{\mathbb{B}}}
\def\bbC{{\mathbb{C}}}
\def\bbD{{\mathbb{D}}}
\def\bbE{{\mathbb{E}}}
\def\bbF{{\mathbb{F}}}
\def\bbG{{\mathbb{G}}}
\def\bbH{{\mathbb{H}}}
\def\bbI{{\mathbb{I}}}
\def\bbJ{{\mathbb{J}}}
\def\bbK{{\mathbb{K}}}
\def\bbL{{\mathbb{L}}}
\def\bbM{{\mathbb{M}}}
\def\bbN{{\mathbb{N}}}
\def\bbO{{\mathbb{O}}}
\def\bbP{{\mathbb{P}}}
\def\bbQ{{\mathbb{Q}}}
\def\bbR{{\mathbb{R}}}
\def\bbS{{\mathbb{S}}}
\def\bbT{{\mathbb{T}}}
\def\bbU{{\mathbb{U}}}
\def\bbV{{\mathbb{V}}}
\def\bbW{{\mathbb{W}}}
\def\bbX{{\mathbb{X}}}
\def\bbY{{\mathbb{Y}}}
\def\bbZ{{\mathbb{Z}}}

\def\cA{\mathcal{A}}
\def\cB{\mathcal{B}}
\def\cC{\mathcal{C}}
\def\cD{\mathcal{D}}
\def\cE{\mathcal{E}}
\def\cF{\mathcal{F}}
\def\cG{\mathcal{G}}
\def\cH{\mathcal{H}}
\def\cI{\mathcal{I}}
\def\cJ{\mathcal{J}}
\def\cK{\mathcal{K}}
\def\cL{\mathcal{L}}
\def\cM{\mathcal{M}}
\def\cN{\mathcal{N}}
\def\cO{\mathcal{O}}
\def\cP{\mathcal{P}}
\def\cQ{\mathcal{Q}}
\def\cR{\mathcal{R}}
\def\cS{\mathcal{S}}
\def\cT{\mathcal{T}}
\def\cU{\mathcal{U}}
\def\cV{\mathcal{V}}
\def\cW{\mathcal{W}}
\def\cX{\mathcal{X}}
\def\cY{\mathcal{Y}}
\def\cZ{\mathcal{Z}}

\def\sfA{\mathsf{A}}
\def\sfB{\mathsf{B}}
\def\sfC{\mathsf{C}}
\def\sfD{\mathsf{D}}
\def\sfE{\mathsf{E}}
\def\sfF{\mathsf{F}}
\def\sfG{\mathsf{G}}
\def\sfH{\mathsf{H}}
\def\sfI{\mathsf{I}}
\def\sfJ{\mathsf{J}}
\def\sfK{\mathsf{K}}
\def\sfL{\mathsf{L}}
\def\sfM{\mathsf{M}}
\def\sfN{\mathsf{N}}
\def\sfO{\mathsf{O}}
\def\sfP{\mathsf{P}}
\def\sfQ{\mathsf{Q}}
\def\sfR{\mathsf{R}}
\def\sfS{\mathsf{S}}
\def\sfT{\mathsf{T}}
\def\sfU{\mathsf{U}}
\def\sfV{\mathsf{V}}
\def\sfW{\mathsf{W}}
\def\sfX{\mathsf{X}}
\def\sfY{\mathsf{Y}}
\def\sfZ{\mathsf{Z}}

\def\bydef{:=}
\def\sfa{{\mathsf{a}}}
\def\sfb{{\mathsf{b}}}
\def\sfc{{\mathsf{c}}}
\def\sfd{{\mathsf{d}}}
\def\sfee{{\mathsf{e}}}
\def\sfff{{\mathsf{f}}}
\def\sfg{{\mathsf{g}}}
\def\sfh{{\mathsf{h}}}
\def\sfi{{\mathsf{i}}}
\def\sfj{{\mathsf{j}}}
\def\sfk{{\mathsf{k}}}
\def\sfl{{\mathsf{l}}}
\def\sfm{{\mathsf{m}}}
\def\sfn{{\mathsf{n}}}
\def\sfo{{\mathsf{o}}}
\def\sfp{{\mathsf{p}}}
\def\sfq{{\mathsf{q}}}
\def\sfr{{\mathsf{r}}}
\def\sfs{{\mathsf{s}}}
\def\sft{{\mathsf{t}}}
\def\sfu{{\mathsf{u}}}
\def\sfv{{\mathsf{v}}}
\def\sfw{{\mathsf{w}}}
\def\sfx{{\mathsf{x}}}
\def\sfy{{\mathsf{y}}}
\def\sfz{{\mathsf{z}}}
\def\sf0{{\mathsf{0}}}

\def\Nt{{N_t}}
\def\Nr{{N_r}}
\def\Ne{{N_e}}
\def\Ns{{N_s}}
\def\Es{{E_s}}
\def\No{{N_o}}
\def\sinc{\mathrm{sinc}}
\def\dmin{d^2_{\mathrm{min}}}
\def\vec{\mathrm{vec}~}
\def\kron{\otimes}
\def\Pe{{P_e}}
\newcommand{\expeq}{\stackrel{.}{=}}
\newcommand{\expg}{\stackrel{.}{\ge}}
\newcommand{\expl}{\stackrel{.}{\le}}
\def\SIR{{\mathsf{SIR}}}

\def\nn{\nonumber}

\maketitle
\vspace{-0.5in}
\begin{abstract}
We consider a node where packets of fixed size (in bits) are generated at arbitrary intervals. The node is required to maintain the peak age of information (AoI) at the monitor below a threshold by transmitting potentially a subset of the generated packets. At any time, depending on the packet availability and the current AoI, the node can choose which packet to transmit, and at what transmission speed (in bits per second). Power consumption is a monotonically increasing convex function of the transmission speed. In this paper, for any given time horizon, the objective is to find a causal policy that minimizes the total energy consumption while satisfying the peak AoI constraint. We consider competitive ratio as the performance metric, that is defined as the ratio of the expected cost of a causal policy, and the expected cost of an optimal offline policy that knows the input (packet generation times) in advance. We first derive a lower bound on the competitive ratio of all causal policies, in terms of the system parameters (such as power function, packet size and peak AoI threshold), and then propose a particular policy for which we show that its competitive ratio has similar order of dependence on the system parameters as the derived lower bound. 
\end{abstract}


\section{Introduction}


In the times of COVID-19 pandemic, real-time networked applications such as autonomous vehicles, immersive gaming, telemedicine and telesurgery played a vital role in enabling people to maintain social-distancing while minimizing the loss of lives and productivity \cite{xiao2018covid}. However, these applications face a serious challenge since they require the data to be fresh at the destination nodes at all times. For this, a typical approach is to quantify the freshness of the data at the destination nodes using a relevant metric, and then optimize that by adopting an appropriate strategy. 

In past, several metrics have been proposed to quantify data-freshness \cite{kosta2017age,maatouk2020age}, most popular being the \emph{age of information} (AoI) \cite{kaul2012status,kaul2012real}. At any time $t$, AoI is equal to the time elapsed since the generation time of the latest data-packet of the transmitting node that has been received at the destination node until time $t$. For data-freshness, low AoI is desirable, and hence, a strategy is sought to minimize AoI.

In practice, AoI minimization is a complicated task. Factors such as limited energy, unavailability of fresh data-packets, etc., often restrict the control options for minimizing AoI.
For example, to maximize the operational life, devices with limited energy need to restrict the number of data-packets that they transmit, as well as their transmission rate (speed), thus leading to large AoI. Hence, there is an inherent AoI-energy tradeoff that any transmission policy must consider \cite{saurav2020game,gu2019timely}. 

In prior work, AoI-energy tradeoff has been considered primarily for $(i)$ energy harvesting model, and $(ii)$ energy conservation model. In energy harvesting model, energy arrives at nodes intermittently, and at any time, depending on the energy level, the expected future energy arrivals, and the packet availability, nodes need to make transmission decisions 
to minimize the AoI \cite{yates2015lazy,arafa2017agedelays,bacinoglu2015age,bacinoglu2017scheduling}. In contrast, for the energy conservation model, there is no limit on the total energy that can be used, and 
the AoI-energy tradeoff is formulated as an optimization problem that seeks 
to minimize either $(i)$ AoI (energy consumption) subject to energy (AoI) constraint \cite{fountoulakis2021joint,tang2020minimizing,moltafet2020power,sun2019sampling}, or $(ii)$ a linear combination of AoI and energy consumption \cite{nived,tseng2019online,saurav2021minimizing,yun2018optimal}. 

In this paper, we focus on the AoI-energy tradeoff for the energy conservation model. For this, we restrict our attention to wired network systems such as smart cars (controller area networks), virtual reality interfaces, motherboards (network of input buffers and central processor), etc., where wireless channel aspects (e.g., fading, interference, etc.) are not applicable. 
The description of the considered system is as follows.
There is a node-monitor pair, linked via a (reliable) wired communication channel.  
Data-packets (in short, packets) of fixed size (say, $W$ bits) are generated at the node at arbitrary time instants. 
The objective of the node is to transmit these packets (possibly only a subset) so that the peak AoI at the monitor is maintained below a threshold at all times. Since each packet contains $W$ bits, the AoI reduces at the monitor at the time instant when the entire $W$ bits of any packet are received.
The node can transmit data to the monitor at a tuneable/variable speed (henceforth, called the \emph{speed scaling model}), by paying a corresponding price in terms of power/energy. 
In particular, if the transmission speed (in bits/sec) is $s(t)$ at time $t$, then the power consumption at time $t$ is given by $P(s(t))$, where $P$ is a non-decreasing convex function. Two popular examples specific to wired systems are $P(s) = s^\alpha$ ($\alpha>1$) and $P(s)=2^s-1$. 

Clearly, there is a tradeoff between power/energy consumption and low AoI, since large transmission speed reduces the AoI, however, at the cost of increased power consumption. Thus, 
the problem we consider is: minimize the total energy used such that the peak AoI at the monitor is below a fixed threshold at all times, over a fixed horizon of time.

One additional feature of the considered problem is that,  at any time, if a new packet is generated, the node may discard the packet currently under transmission, and switch to transmitting the newly generated packet. This action leads to a larger reduction in the AoI at the monitor, however, at the cost of wasting energy already consumed for partially transmitting the discarded packet.  
Thus, the decisions to make for an algorithm at each time are : whether to transmit an available packet, discard the ongoing packet transmission on generation of a new packet, and the speed at which to transmit the chosen packet at any time.


\subsection{Prior Work}

In past, AoI-energy tradeoff has been considered extensively \cite{saurav2020game,gu2019timely,yates2015lazy,arafa2017agedelays,bacinoglu2015age,bacinoglu2017scheduling,fountoulakis2020optimal,tang2020minimizing,moltafet2020power,nived,tseng2019online,saurav2021minimizing,yun2018optimal,bhat2020throughput,bhat2021minimization}. However, with speed scaling model, the problem is relatively unexplored, and only a few works in the past have considered it \cite{bhat2020throughput,bhat2021minimization,arafa2017agedelays}. 
For example, \cite{bhat2020throughput,bhat2021minimization} considered a throughput maximization problem over a fading channel, where 
the number of bits transmitted (throughput) in each slot is proportional to the fading gain and the transmitted power. For each slot, if the number of transmitted bits is above a fixed threshold, the AoI is reset to one, otherwise it is incremented by one. Enforcing two constraints, an upper bound on the average AoI and an average power constraint, \cite{bhat2020throughput,bhat2021minimization} proposes an algorithm that is shown to be $2$-approximate. 
For a similar fading model with a multiple access channel, \cite{bhat2021minimization} considered the problem of minimizing average AoI under an average power constraint.
An alternate speed scaling formulation has been considered in  \cite{arafa2017agedelays}, where the delay incurred in transmitting an update is controlled by the amount of energy used, and the problem is to minimize AoI subject to the energy neutrality constraint enforced by the energy harvesting model.

The problem considered in this paper, to minimize energy subject to peak AoI constraint is similar to the classical 
energy minimization problem of scheduling jobs with deadlines (say, \textsc{DeadlineSched}) \cite{yao1995scheduling,bansal2007speed}. With \textsc{DeadlineSched}, there is a single server that can process jobs at speed $s$ with power cost $P(s)$. Jobs with associated deadlines arrive  sequentially, 
and the server needs to process all the arriving jobs before their corresponding deadlines, while consuming minimum energy.  
When power function is $P(s)=s^\alpha$ ($\alpha>1$), \cite{yao1995scheduling} proposed a causal preemptive policy for which the competitive ratio (i.e., the ratio of the energy consumed by the causal policy to the energy consumed by an optimal offline policy, maximized over all inputs) is at most $\alpha^\alpha$ \cite{bansal2007speed}.  
In the special case when the power function is $P(s)=2^s-1$, and all the packets are of equal size, with a common deadline, \cite{deshmukh2016online,rahman2020deadline} gave a $3$-competitive policy for \textsc{DeadlineSched}. 

The similarity between \textsc{DeadlineSched} and the considered problem stems from the fact that at each time instant, the peak AoI constraint imposes a deadline before which a 
new packet must be transmitted. However, the key difference between the two problems is that with \textsc{DeadlineSched} every job that arrives  needs to be processed completely, while in the considered problem, there is only a peak AoI constraint which could be satisfied by transmitting only a subset of the generated packets. Since the objective is to minimize total energy, subject to peak AoI constraint, an additional challenge compared to \cite{yao1995scheduling,bansal2007speed} (where only speed and the order of processing is to be decided) is to identify the subset of packets to transmit.





In this paper, we consider a general setting, 
where the time horizon is finite, the packet generation times are arbitrary, and there is a peak AoI constraint.  
To put this into perspective, we briefly discuss relevant prior work as follows.
\paragraph{Finite Time Horizon} Most of the prior work pertaining to the AoI minimization problem considers infinite time horizon \cite{saurav2020game,kaul2012status,kaul2012real,gu2019timely,bhat2020throughput,moltafet2020power,yates2015lazy,nived,saurav2021minimizing}. Infinite time horizon assumption makes it possible to utilize stochastic properties of the problem, such as distribution over the packet inter-generation times/energy arrival instants, etc. This also makes it possible to ignore the edge effects (such as initial AoI, cost/energy incurred at times close to $0$ or time horizon $T$, etc.). However, in this paper, we consider finite time horizon (such as in \cite{bastopcu2019minimizing,tseng2019online}), and fully acknowledge the edge effects.

\paragraph{Arbitrary Packet Inter-generation Times}
In prior work, packet inter-generation time patterns have been considered for the following three models: $(i)$ generate-at-will model --- a fresh packet is available at the node at all times (e.g., \cite{saurav2020game,bhat2020throughput,moltafet2020power,tseng2019online,sun2017update}), $(ii)$ stochastic arrival model --- fresh packets are generated with inter-generation time following particular distribution (e.g., \cite{kaul2012real,gu2019timely,saurav2021minimizing,kadota2021age,kadota2019minimizing}), and $(iii)$ arbitrary arrival model --- packet inter-generation times are arbitrary, and hence, need not follow any particular distribution or pattern (e.g., \cite{nived,sinha2020optimizing,bedewy2016optimizing,he2016optimizing,sun2018age}). Clearly, arbitrary arrival model is the most general one, and may even capture adversarial inputs. 

\paragraph{Peak AoI Constraint}
Primarily, in AoI-related problems, the actual metric being minimized or considered is either $(i)$ time-averaged AoI $\Delta_{av}(T)=\frac{1}{T}\int_{t=0}^{T}\Delta(t)dt$ (where $T$ is the time horizon, while $\Delta(t)$ denotes instantaneous AoI at time $t$) \cite{saurav2021minimizing,saurav2020game,moltafet2020power,kaul2012real,kaul2012status,bhat2020throughput,bhat2021minimization}, or $(ii)$ average of Peak AoI $\Delta_P(T)=\frac{1}{\cN(T)}\sum_{i=1}^{\cN(T)}\Delta_p(i)$ (where $\cN(T)$ denotes the number of packets successful transmitted by the node until time horizon $T$, while $\Delta_p(i)$ denotes the AoI just before the completion of $i^{th}$ transmission) \cite{costa2014age,talak2019optimizing,elmagid2019average,khorsandmanesh2021average}. However, in practice these metrics (that consider only average values), may not be sufficient for time-critical applications such as telesurgery, driverless car, etc. For such applications, 
peak AoI constraint \cite{li2020aoi,ostman2019peak,devassy2019reliable} is a more suitable choice as it guarantees that the instantaneous AoI is below a given threshold at all times. 
Moreover, as we show in Appendix \ref{app:AAoI} (in the supplementary material), for the considered model with arbitrary packet arrival and variable speed, any causal policy that seeks to minimize energy consumption under an average AoI constraint, has unbounded competitive ratio. Therefore, average AoI constraints are considered mostly for simpler models with stochastic arrival, fixed transmission speed or infinite time horizon  \cite{bhat2020throughput,fountoulakis2021joint}.

\subsection{Our Contributions}
The main contributions of this paper are as follows. 
	
	1) When all the generated packets are of same size ($W$ bits), we derive a lower bound on the competitive ratio of all causal policies $\pi$ that seek to minimize energy consumption while satisfying a peak AoI constraint at all time instants. We show that for convex power function $P(s(t))$ (where $s(t)$ denotes the transmission speed at time $t$), the competitive ratio of any causal policy $\pi$ is 
	\begin{align} \label{intro:eq:cr-lb}
		\textsc{cr}_{\pi}\ge c_1P(c_2\hat{s})/P(c_3\hat{s}),
	\end{align}
	where $\hat{s} = W/D$ ($W$ is the packet size, and $D$ is the peak AoI constraint), while $c_1$, $c_2$ and $c_3$ are finite positive constants such that $c_1>0.13$, $c_2-c_3\ge 0.14$ and $c_2/c_3\ge 1.07$. Thus for any causal policy $\pi$, $(i)$ if $P(s)=2^s-1$, the competitive ratio $\textsc{cr}_{\pi}\ge 0.13(2^{0.14\hat{s}})$, i.e. it increases exponentially with increase in $\hat{s}$,  
	while $(ii)$ if $P(s)=s^\alpha$ ($\alpha>1$), the competitive ratio $\textsc{cr}_{\pi}\ge 0.13(1.07^\alpha)$, i.e. it increases exponentially with increase in $\alpha$. 	
	
	
	2) To design a policy for the considered problem, we need to specify, at any time  
	i) which packet to transmit from the set of available packets (may require preempting an ongoing transmission), and ii) the transmission speed. 
	Our proposed causal policy $\pi^g$ does the following. Assuming that the sizes of all generated packets is $W$ bits, 
	at time $t$ when no packet is under transmission, $\pi^g$ begins to transmit the latest generated packet with constant speed $s^g(t)=\max\{3W/D,W/(D-\Delta(t))\}$, where $\Delta(t)$ denotes the instantaneous AoI at time $t$. The policy $\pi^g$ never preempts an ongoing update transmission. 
	Using the specific choice of speed $s^g(t)$, we show that for convex power functions $P(\cdot)$, the competitive ratio of $\pi^g$ is 
	\begin{align} \label{intro:eq:cr-ub}
		\textsc{cr}_{\pi^g}\le 2P(3\hat{s})/P(\hat{s})+1,
	\end{align}  
	where recall that $\hat{s}=W/D$. Consequently, we get that $(i)$ for $P(s)=s^\alpha$ ($\alpha>1$), $\textsc{cr}_{\pi^g}\le 2\cdot3^\alpha+1$, 
	while $(ii)$ for $P(s)=2^s-1$, $\textsc{cr}_{\pi^g}\le 2(2^{3\hat{s}}-1)/(2^{\hat{s}}-1)+1$.
	Comparing \eqref{intro:eq:cr-lb} and \eqref{intro:eq:cr-ub}, we can conclude that any other policy (that could possibly be preemptive) can improve the competitive ratio of $\pi^g$ by at most a constant.
	
	3) We also consider the problem in a more practical case, where at each time instant, there is a limit on the peak transmit power (speed). In particular, let $s^{\max}$ be the maximum speed at which the node may transmit at any time, and $\Sigma(s^{\max})$ be the set of packet generation sequences $\sigma$ for which the peak AoI constraint can be satisfied under some causal policy. Our results can be broken down in the following three cases :
	
	$(i)$ When $s^{\max}\ge 3\hat{s}$: In this case, we consider the policy $\pi^g$ defined earlier for the case when there is no upper bound on the transmit power at any time with no modification. 
	Using the structure of the packet generation sequence set $\Sigma(s^{\max})$,
	we show that for any 
	$\sigma\in\Sigma(s^{\max})$, $\pi^g$ never needs to choose a speed greater than $s^{\max}$. Because of this automatic satisfaction of the peak transmit power constraint by $\pi^g$, we also get that $\pi^g$ satisfies the peak AoI constraint 
	and has competitive ratio (defined with respect to the set 
	$\Sigma(s^{\max})$) equal to
	$\textsc{cr}_{\pi^g}\le 2P(3\hat{s})/P(\hat{s})+1$. 
	$(ii)$ When $s^{\max}\in[2\hat{s},3\hat{s})$: We show that for any causal policy $\pi$, there exists some packet generation sequence $\sigma\in\Sigma(s^{\max})$ for which the peak AoI constraint is violated. Hence, the competitive ratio is unbounded for all causal policies. $(iii)$ When $s^{\max}<2\hat{s}$:  We show that the peak AoI constraint can be satisfied only when the maximum allowed AoI $D>2T/3$, where $T$ denotes the time horizon, and derive an optimal causal policy for this case.
	
	Thus, 
	we observe two conflicting phenomena. 
	i) When $s^{\max}$ is small, packets need to be transmitted at low speed over long time intervals, and there is lesser time to adapt to different possible future packet arrivals. Therefore, when $s^{\max}\in[2\hat{s},3\hat{s})$, no causal policy can satisfy the peak AoI constraint for all sequences $\sigma\in\Sigma(s^{\max})$.
	ii) As $s^{\max}$ decreases, the size of the set $\Sigma(s^{\max})$ (number of packet generation sequences for which the peak AoI constraint can be satisfied) decreases. When $s^{\max}<2\hat{s}$, the remaining sequences $\sigma\in\Sigma(s^{\max})$ have similar patterns, and there exists an optimal causal policy. 
	
	4) When the generated packets are of arbitrary sizes in the range $[w,W]$ bits, and there is no peak transmit power constraint, we show that the competitive ratio of the proposed policy $\pi^g$ that uses $W$ for defining the speed is $\textsc{cr}_{\pi^g}\le 2P(3\zeta\tilde{s})/P(\tilde{s})+1$, where $\zeta=W/w$, and $\tilde{s}=w/D$. Moreover, if $\zeta$ is unbounded, the competitive ratio of any causal policy $\pi$ is unbounded.

\section{System Model} \label{sec:SystemModel}
Consider a node where data-packets (in short, packets), each of size $W$ bits are generated intermittently. In particular, the $i^{th}$ packet at the node is generated at time $t_i$, where $t_i$ is determined by external factors (possibly adversarial), with inter-generation time $X_i=t_i-t_{i-1}$.  
A packet $i$ is said to be delivered (at the monitor) at time $\tau_i$ if the node finishes transmitting the $W$ bits of packet $i$ at time $\tau_i$. 

At any time $t$, the \emph{age of information} (AoI) at the monitor is equal to $\Delta(t)=t-\mu(t)$, where $\mu(t)$ is the generation time of the latest packet that has been delivered to the monitor until time $t$. Further, in an interval $[0,T]$, peak AoI at the monitor is defined to be $\max_{t\in[0,T]}\Delta(t)$.
We consider a peak AoI constraint, i.e., for any given time horizon $T$, the node requires that the peak AoI at the monitor in the interval $[0,T]$ is less than $D$ (where $D$ is a known constant), as shown in Figure \ref{fig:age}. 
\begin{assumption} \label{assume:epsilon}
	For the problem to be meaningful, we assume that the packet generation times, and the initial AoI $\Delta(0)$ are such that there exists some causal policy that can satisfy the peak AoI constraint.
\end{assumption}
\begin{remark} \label{remark:X<D}
	Note that if the packet inter-generation time $X_i\ge D$ for any $i$, then the peak AoI constraint can never be satisfied. 
	Therefore, Assumption \ref{assume:epsilon} implies that 
	$X_i<D$, $\forall i$. 
\end{remark}
\begin{figure} 
	\begin{center}
		\begin{tikzpicture}[thick,scale=1, every node/.style={scale=1}]
		\draw[->] (-0.25,0) to (7,0) node[below]{time ($t$)};
		\draw[->] (0.35,-0.25) node[left]{$0$} to (0.35,3.5) node[left]{$\Delta(t)$};
		\draw[dashed] (0.35,3) node[left]{$D$} to (7,3);
		\draw (0.35,0.65) node[left]{$\Delta(0)$} to (2,2.3) to (2,1) to (4,3) to (4,1) to (5.5,2.5) to (5.5,0.75) to (6.25,1.5); 
		\draw[dashed] (2,2.3) to (2.7,3) to (2.7,0);
		\draw[dashed] (5.5,2.5) to (6,3) to (6,0);
		
		\draw[|->|] (0.34,-0.8) -- (2.7,-0.8) node[rectangle,inner sep=-1pt,midway,fill=white]{$d_0$};
		
		\draw[|<->|] (1,-1.15) -- (4,-1.15) node[rectangle,inner sep=-1pt,midway,fill=white]{$D$};
		
		\draw[|<->|] (3,-0.8) -- (6,-0.8) node[rectangle,inner sep=-1pt,midway,fill=white]{$D$};
		
		\draw[loosely dotted] (6.5,1) to (7,1); 
		
		\draw (1,-0.1) node[below]{$t_1$} to (1,0.1);
		\draw (2,-0.1) to (2,0.1);
		\draw (2,-0.1) node[below]{$\tau_1$};
		\draw (2.7,-0.1) to (2.7,0.1);
		\draw (2.6,-0.4) node{$d_0$};
		\draw (3,-0.1) node[below]{$t_2$} to (3,0.1);
		
		\draw (4,-0.1) node[below]{$\tau_2, d_1$} to (4,0.1);
		\draw (4.75,-0.1) node[below]{$t_3$} to (4.75,0.1);
		\draw (5.5,-0.1) to (5.5,0.1); 
		\draw (5.5,-0.1) node[below]{$\tau_3$};
		\draw (6,-0.1) node[below]{$d_2$} to (6,0.1);
		
		\draw[dashed] (1,0) to (2,1);
		\draw[dashed] (3,0) to (4,1);
		\draw[dashed] (4.75,0) to (5.5,0.75);
		
		\draw[dashed] (2,0.1) to (2,2.3);
		\draw[dashed] (4,0.1) to (4,2.3);
		\draw[dashed] (5.5,0.1) to (5.5,2.3);
		
		
		
		
		\end{tikzpicture}
		\caption{A typical AoI profile of node. Here, $d_i=t_i+D$.} 
		\label{fig:age} 
	\end{center}
\end{figure}
\begin{remark}
	At any time $t$, the peak AoI constraint only requires that $\mu(t)$, i.e., the generation time of the latest packet delivered to the monitor until time $t$ is less than $D$. 
	Therefore, to satisfy the peak AoI constraint, it is not necessary to transmit every generated packet. For example, at time $t$, if packets $i$ and $j$ (with generation times $t_i$ and $t_j$ respectively) are available at the node (i.e., neither of the packets have been completely transmitted until time $t$), and $t_i<t_j$, then for satisfying the peak AoI constraint, it is sufficient to transmit packet $j$ only, without transmitting packet $i$. Moreover, after packet $j$ gets delivered to the monitor, transmitting packet $i$ is no longer useful (and hence, need not be transmitted) because the peak AoI will still be determined by $t_j$, i.e., the generation time of the latest packet delivered at the monitor. 
\end{remark}

We consider a speed scaling model, where at any time $t$, the node can transmit a packet at speed $s(t)\ge0$ (in bits/sec) adaptively, i.e., the node can choose $s(t)$ using causal information available at time $t$. 
Also, transmitting a packet at speed $s(t)$ consumes power
$P(s(t))$, which is an increasing and convex function of speed $s(t)$, e.g., $P(s)=s^\alpha$ ($\alpha>1$), or $P(s)=2^s-1$, motivated by Shannon's rate function. 
Therefore, if the node transmits packets at high speed, it incurs low AoI, but consumes large amount of energy. 
Hence, finding an optimal transmission speed at each time $t$ is a non-trivial task.
\begin{assumption} \label{assume:P0}
	Throughout this paper, we assume that when speed $s(t)=0$, the power consumption is $P(0)=0$. This assumption does not affect the results derived in this paper, but allows us to ignore the terms containing $P(0)$ which merely appear as an offset.
\end{assumption}

\begin{definition} 
\label{def:preemptive-policy}
A policy that at any time $t$, \emph{can} interrupt (stop) an ongoing transmission of a packet, and begin transmitting a newly generated packet, is called an interruptive policy.   
When a transmission is interrupted, the packet that was being transmitted 
is discarded, since if a newly generated packet is delivered at the monitor, all previously generated packets become useless in minimizing AoI.\footnote{An interruptive policy, unlike preemptive policy, discards packets.}   
\end{definition}
\begin{remark}
	Note that the class of interruptive policies, by definition, includes all non-interruptive policies, that never interrupts transmission of any packet. Thus, the set of all causal policies implicitly refer to the set of all interruptive causal policies that may interrupt ongoing packet transmission. 
\end{remark}

In this paper, we consider the problem of finding an interruptive causal policy (in short, causal policy) that chooses the packets to transmit, the time interval over which the packets are transmitted, and their instantaneous transmission speed $s(t)$, so that 
the peak AoI is maintained below $D$ at all times over a time horizon $T$, while consuming minimum energy. Formally, 
the objective can be stated as follows.
 \begin{subequations}
	\begin{align} \label{eq:objective-init}
		&\underset{\pi\in\Pi}{\min}\ \ E_{\pi}(\sigma) = \int_{t=0}^{T}P(s(t))dt \\
		\label{eq:constraint}
		&\text{s.t.}\ \  \Delta(t)< D, \ \ \forall t\in [0,T], 
	\end{align}
\end{subequations}
where $\Pi$ is the set of all causal policies\footnote{Although packet generation instants are not known in advance, we assume that the time horizon $T$ is known.} for packet scheduling and the choice of speed $s(t)$, and $\sigma=\{t_1,t_2,...\}$ is the sequence of packet generation times. 
Note that the choice of packet a policy transmits at any time $t$ is inherently captured by \eqref{eq:objective-init}.

\begin{remark} \label{remark:power-constraint} 
	For ease of exposition, the more practical case for studying problem \eqref{eq:objective-init}--\eqref{eq:constraint} when there is a limit on the transmit power/maximum speed at any time instant is studied in Section \ref{sec:power-constraint}. 
\end{remark}

\begin{definition} \label{def:offline-optimal-policy}
	A policy $\pi^\star$ is said to be offline optimal if it satisfies the peak AoI constraint \eqref{eq:constraint}, and there exists no other policy $\tilde{\pi}$ that can simultaneously satisfy the peak AoI constraint \eqref{eq:constraint}, and consume less energy than $\pi^\star$, even if $\tilde{\pi}$ knows the generation time of all the packets in advance. 
	Optimal offline policies are useful as they provide a lower bound on the energy consumed by any causal policy $\pi$.
\end{definition}

From prior work \cite{yao1995scheduling,deshmukh2016online,rahman2020deadline}, it is known that finding an optimal causal policy for energy minimization problems under hard constraints (such as individual/common deadline for packets) is a challenging task. Hence, a usual approach is to 
find a causal policy $\pi$ whose competitive ratio, defined as the ratio of the energy consumed by a causal policy $\pi$ (to satisfy the peak AoI constraint) and the energy consumed by an optimal offline policy $\pi^\star$ (Definition \ref{def:offline-optimal-policy}), maximized over all possible sequence of packet generation times $\sigma$, is small. Mathematically, the competitive ratio of policy $\pi$ is 
\begin{align} \label{eq:CR-definition}
	\textsc{cr}_\pi = \max_{\sigma}\frac{E_\pi(\sigma)}{E_{\pi^\star}(\sigma)}.
\end{align}
By definition, a policy with small competitive ratio performs well for all sequences of packet generation times $\sigma$, and hence, is robust.
In the rest of the paper, we will consider a particular non-interruptive causal policy (Definition \ref{def:preemptive-policy}), and show that its competitive ratio is at most $2P(3\hat{s})/P(\hat{s})+1$, where $\hat{s}=W/D$. We will also derive a general lower bound on the competitive ratio of all causal policies, and show that for different power functions of interest, the competitive ratio of the considered policy has similar characteristics (dependence on parameters) as the derived lower bound, which is policy independent. 
Note that for a non-interruptive causal policy which is much simpler to implement than a general (interruptive) causal policy, this is a significant result.  

 \subsection{An Equivalent Deadline Constraint Problem} \label{sec:equivalent-interpretation}
 The peak AoI constraint \eqref{eq:constraint} can also be interpreted as a deadline constraint, where a deadline is defined as follows.
 \begin{definition} \label{def:deadline}
 	At any time $t$, the deadline $d(t)$ is defined as the earliest time instant at which the peak AoI constraint \eqref{eq:constraint} will be violated if no packet is delivered to the monitor after time $t$. Thus, $d(t)=t+(D-\Delta(t))=\mu(t)+D$, where $\mu(t)=t-\Delta(t)$ is the generation time of the latest packet that has been delivered to the monitor until time $t$. 
 \end{definition} 
\begin{remark} \label{remark:discontinuous-jump}
	Note that $d(t)=\mu(t)+D$ is a non-decreasing function of $t$. In fact, deadline $d(t)$ increases in steps whenever a fresh packet $j$ ($t_j>\mu(t)$; see Definition \ref{def:fresh} below) is delivered to the monitor. This happens because $\mu(t)$ (i.e., the generation time of the latest packet delivered until time $t$) increases discontinuously to the generation time $t_j$ of packet $j$, at the instant packet $j$ is delivered to the monitor. 
\end{remark}
\begin{definition} \label{def:fresh}
	At any time $t$, a packet $i$ is defined to be fresh if its generation time $t_i\le t$ is greater than $\mu(t)$, i.e. $t_i > \mu(t)$. Otherwise, the packet is stale.
\end{definition}

Note that the peak AoI constraint \eqref{eq:constraint} is satisfied if and only if $D-\Delta(t)>0$, for all $t\in[0,T]$. Also, at any time $t$, $d(t)=t+(D-\Delta(t))$ is greater than $t$ if and only if $D-\Delta(t)>0$. Therefore, the peak AoI constraint \eqref{eq:constraint} is equivalent to the following deadline constraint:
		\begin{align} \label{eq:deadline-constraint-init}
			d(t)> t, \ \ \forall t \in [0,T].
		\end{align} 
	In other words, the peak AoI constraint \eqref{eq:constraint} is equivalent to the constraint that at any time $t\in[0,T]$, the current deadline $d(t)$ must be in future. Hence, instead of \eqref{eq:objective-init}--\eqref{eq:constraint}, hereafter we consider the following equivalent optimization problem.
	 \begin{subequations}
		\begin{align} \label{eq:objective}
			&\underset{\pi\in\Pi}{\min}\ \ E_{\pi}(\sigma) = \int_{t=0}^{T}P(s(t))dt \\
			\label{eq:deadline-constraint}
			&\text{s.t.}\ \  d(t)> t, \ \ \forall t\in [0,T]. 
		\end{align}
	\end{subequations}
	\begin{remark}
		As we show in Section \ref{sec:opt-policy}, considering peak AoI constraint \eqref{eq:constraint} as deadline constraint \eqref{eq:deadline-constraint} reveals several key properties of an optimal offline policy $\pi^\star$, and simplifies the overall analysis in this paper.
	\end{remark} 

	\begin{definition} \label{def:feasible}
		A policy $\pi$ is defined to be feasible if it satisfies the deadline constraint \eqref{eq:deadline-constraint} at all times $t\in[0,T]$.   	
	\end{definition}

	 \begin{proposition} \label{prop:feasible}
		At any time $t\in[0,T]$, if $d(t)\le T$, then for any feasible policy $\pi$ (Definition \ref{def:feasible}), at least one fresh packet has to be delivered to the monitor in interval $[t,d(t))$.
%
	\end{proposition}
    \begin{IEEEproof}
    	If the deadline at time $t\in[0,T]$ is $d(t)\le T$, and no fresh packet is delivered to the monitor in interval $[t,d(t))$, then the deadline constraint \eqref{eq:deadline-constraint} will be violated at time $d(t)$ (follows from the definition of $d(t)$; Definition \ref{def:deadline}). 
    	Hence, a feasible policy must deliver at least one fresh packet to the monitor in interval $[t,d(t))$, for all $t\in[0,T]$, if $d(t)\le T$.
    \end{IEEEproof}

 \begin{remark} \label{remark:only-fresh-packet}
	In order to satisfy the deadline constraint \eqref{eq:deadline-constraint}, note that only fresh packets are needed/useful. Therefore, if a policy transmits a stale packet, it wastes energy. Hence, in the rest of this paper, we only consider packets that are fresh, and at any time, the term `packet' implicitly means a fresh packet. 
\end{remark}

\begin{definition} \label{def:d_i}
	For each packet $i$ generated at time $t_i$, we define $d_i=t_i+D$. 
\end{definition}

\subsection{Property of Convex Power Function} \label{sec:convexity}

Convexity of power function implies the following property.

\begin{lemma} \label{lemma:constant-speed-better}
	Energy consumed in transmitting $w$ bits in a fixed interval $[p,q)$ is minimum if the bits are transmitted at a constant speed $s_w(t)=w/(q-p)$. Also, the minimum energy consumed in interval $[p,q)$ is $P(w/(q-p))(q-p)$.
\end{lemma}

\begin{corollary} \label{cor:P(w/y)y-decreasing-y}
	For fixed $w$, $P(w/y)y$ decreases with increase in $y$.
\end{corollary}

For the proofs of Lemma \ref{lemma:constant-speed-better} and Corollary \ref{cor:P(w/y)y-decreasing-y}, please refer to Appendix \ref{appendix:constant-speed-better} and \ref{appendix:proof-cor:P(w/y)y-decreasing-y}, respectively.
	\textbf{All the appendices referred in this paper are available in the supplementary material}.


\section{Limitations of a Causal Policy $\pi$} \label{sec:limitationsCausalPolicy}
Before we discuss a particular causal policy for minimizing the energy consumption \eqref{eq:objective} (under the deadline constraint \eqref{eq:deadline-constraint}), it is important to note the fundamental limitations of any causal policy $\pi$. 
Towards that end, Theorem \ref{thm:lb-CR} provides a lower bound on the competitive ratio of any causal policy $\pi$, and shows (in Corollary \ref{cor:lb-CR}) that for certain power functions $P(\cdot)$, the competitive ratio of any causal policy $\pi$ is an increasing function of $W/D$.
\begin{theorem} \label{thm:lb-CR}
	For any causal policy $\pi$, its competitive ratio
	\begin{align} \label{eq:lb-CR}
		\textsc{cr}_{\pi}\ge \frac{c_1P(c_2W/D)}{P(c_3W/D)},
	\end{align}
	where $c_1$, $c_2$ and $c_3$ are finite positive constants, $c_1>0.13$, $c_2-c_3\ge 0.14$, and $c_2/c_3\ge 1.07$.
\end{theorem} 
\begin{IEEEproof}[Proof Sketch]
	To prove Theorem \ref{thm:lb-CR}, we consider a particular scenario where the AoI at the monitor at time $t=0$ is $\Delta(0)=D/2$, the time horizon $T=3D/2-\delta$ (for $\delta\to 0^+$), and the packets are generated according to one of the two instances of packet generation times $\sigma$: $(i)$ $\sigma_1=\{0,D/4,D/2\}$, and $(ii)$ $\sigma_2=\{0,D/4,5D/6\}$. Then, for indexing all causal policies, we consider different cases based on the packet(s) that any causal policy may transmit 
	in interval $[0,D/2)$, and for each case, we compute the competitive ratio \eqref{eq:CR-definition}, where the maximization is with respect to $\sigma\in\{\sigma_1,\sigma_2\}$. Finally, we take the minimum over the competitive ratio obtained for different cases considered above, and obtain \eqref{eq:lb-CR}, where $c_1$, $c_2$ and $c_3$ are finite positive constants, $c_1>0.13$, $c_2-c_3\ge 0.14$, and $c_2/c_3\ge 1.07$. For detailed proof, see Appendix \ref{appendix:proof-thm-lb-CR}.
\end{IEEEproof}
\begin{corollary} \label{cor:lb-CR}
	For any causal policy $\pi$, $(i)$ if $P(s)=s^\alpha$ ($\alpha>1$), the competitive ratio $\textsc{cr}_{\pi}\ge c_1(c_2/c_3)^\alpha \ge 0.13( 1.07^\alpha$) increases exponentially with increase in $\alpha$,  
	while $(ii)$ for $P(s)=2^s-1$, the competitive ratio $\textsc{cr}_{\pi}\ge c_1 2^{(c_2-c_3)W/D}\ge 0.13( 2^{0.14W/D})$ increases exponentially with increase in $W/D$. 
\end{corollary}
\begin{remark}
	In this paper, we have modeled the AoI requirements of the system as peak AoI constraint \eqref{eq:constraint}. In literature (e.g. \cite{bhat2020throughput}), an alternate formulation have been in terms of time-averaged AoI (in short, average AoI) constraint. 
	However, as we have shown in Appendix \ref{app:AAoI}, in the considered setting with arbitrary packet generation times, if the peak AoI constraint \eqref{eq:constraint} is replaced by average AoI constraint, the competitive ratio of all causal policies will be infinite. 
\end{remark}



In the next section, we propose a feasible non-interruptive (causal) greedy policy $\pi^g$, and show that the competitive ratio \eqref{eq:CR-definition} of $\pi^g$ is upper bounded by $c_1'P(c_2'W/D)/P(c_3'W/D)+1$ (where $c_1'=2$, $c_2'=3$, and $c_3'=1$). Thus, we show that the dependence of the competitive ratio of $\pi^g$ on the system parameters is similar to the policy-independent lower bound \eqref{eq:lb-CR} in Theorem \ref{thm:lb-CR}.

\section{A Greedy Policy $\pi^g$} \label{sec:greedy-policy}

Consider a greedy policy $\pi^g$ (Algorithm \ref{algo:greedy}) that at any time $t$, if the node is idle (i.e., not transmitting any packet) and the deadline $d(t)\le T$, transmits the latest available (fresh) packet with \emph{constant} speed $s^g(t)$ \eqref{eq:speed}, starting at time $t$, throughout until the $W$ bits of the packet are delivered to the monitor (and waits otherwise), 
 \begin{align} \label{eq:speed}
	s^{g}(t)=\max\left\{\frac{W}{d(t)-t},\frac{W}{D/3}\right\}. 
\end{align}

\begin{remark} \label{remark:non-preemptive}
	Note that the greedy policy $\pi^g$ 
	computes the speed $s^g(t)$ \eqref{eq:speed} at the instant $t$ when it begins to transmit a packet $i$. Afterwards, the speed remains constant until the $W$ bits of packet $i$ gets transmitted. Also, $\pi^g$
	never interrupts any ongoing packet transmission. However, these are not constraints, and in general, a policy can vary speed while transmitting a packet, and interrupt ongoing packet transmission to solve \eqref{eq:objective}.
\end{remark}

\begin{algorithm}
	\caption{Greedy Policy $\pi^g$.}
	\label{algo:greedy}
	\begin{algorithmic}
		\STATE $t\leftarrow$ current time;
		\IF{$d(t)\le T$ \AND node idle \AND packet available}
		\STATE transmit the latest generated packet with constant speed $s^{g}(t)$ \eqref{eq:speed} throughout the time interval $[t,t+W/s^g(t))$. 
		\ENDIF
	\end{algorithmic}
\end{algorithm} 

Although greedy policy $\pi^g$ appears obvious, the speed $s^g(t)$ \eqref{eq:speed} has been chosen carefully such that $(i)$ $\pi^g$ is feasible ($\pi^g$ never idles when there is a fresh packet to transmit, and since $s^g(t)\ge W/(d(t)-t)$, if $\pi^g$ begins to transmit a packet at time $t$, then the packet will be delivered to the monitor before deadline $d(t)$), 
and $(ii)$ speed $s^g(t)$ cannot be arbitrarily large (in fact, $s^g(t)$ cannot be greater than $3W/D$), unless a particular event happens (defined in Proposition \ref{prop:large-X}) with regard to the packet generation time and for which we can lower bound the energy consumed by an optimal offline policy $\pi^\star$ (Lemma \ref{lemma:partitionAB} in Appendix \ref{appendix:lemma-partitionAB}). 
\begin{proposition} \label{prop:large-X}
	If $\pi^g$ begins to transmit a packet $j$ at time $t$, then the speed $s^g(t)>3W/D$ only if no packet was generated in interval $[t-2D/3,t)$. 
\end{proposition}
\begin{IEEEproof}
	See Appendix \ref{appendix:prop-large-X}.
\end{IEEEproof}
\begin{corollary} \label{cor:large-speed-immediate-transmission}
	If $\pi^g$ transmits a packet $j$ with speed greater than $3W/D$, then it must have begun to transmit packet $j$ immediately after it was generated at time $t_j$, and completed the transmission at time $d(t_j)$.
\end{corollary}
\begin{IEEEproof}
	See Appendix \ref{appendix:proof-cor-large-speed-immediate-transmission}.
\end{IEEEproof}

Next, in Example \ref{ex:bad-variant-greedy}, we illustrate the significance of the term $W/(D/3)$ inside the max function in \eqref{eq:speed}.  
\begin{example} \label{ex:bad-variant-greedy}
	Let at $t=0$, $\Delta(0)=0$, $D=2$, and $T=3+\delta/2$, where $\delta\to 0^+$. Also, let three packets be generated at time $t=0, t=\delta$, and $t=1+\delta$, respectively. Then, an optimal offline policy $\pi^{\star}$ will only transmit the third packet, with constant speed $W/(1-\delta)$ for $(1-\delta)$ time units, whereas the greedy policy $\pi^g$ (Algorithm \ref{algo:greedy}) 
	will transmit all three packets with constant speed $3W/D$ (each for $D/3$ time units). 
	However, if the term $W/(D/3)$ inside the max function in $s^g(t)$ \eqref{eq:speed} is dropped, or replaced by a smaller value, say $2W/D$, 
	then $\pi^g$ will still transmit all three packets in the given scenario, but the third packet will be transmitted with a constant speed $W/\delta\to\infty$, thus consuming infinite amount of energy. 
\end{example}
Essentially, Example \ref{ex:bad-variant-greedy} shows that if the term $W/(D/3)$ is not there in \eqref{eq:speed} (i.e., if the speed $s^g(t)$ is \emph{not} lower bounded by $3W/D$), then for certain inputs $\sigma$ (sequence of packet generation times), $\pi^g$ may spend too much time transmitting a single packet, and hence, may have little time left to transmit the next packet. This would force $\pi^g$ to transmit the next packet at very high speed, thus consuming large amount of energy.

The main result of this section is as follows.
\begin{theorem} \label{thm:CR-bounds}
	Let $\hat{s}=W/D$. The competitive ratio ($\textsc{cr}_{\pi^g}$) of greedy policy $\pi^g$ is bounded as 
	\begin{align} \label{eq:CR-bounds}
		\frac{1.5P(3\hat{s})}{P(1.5\hat{s})}\le \textsc{cr}_{\pi^g}\le \frac{2P(3\hat{s})}{P(\hat{s})}+1.
	\end{align}
\end{theorem}
\begin{remark} \label{remark:cr-poly-exp}
	In \eqref{eq:CR-bounds}, if $P(s)=s^{\alpha}$ ($\alpha>1$), $\textsc{cr}_{\pi^g}\le 2\cdot 3^\alpha+1$, while if $P(s)=2^s-1$, $\textsc{cr}_{\pi^g}\le 2(2^{3W/D}-1)/(2^{W/D}-1)+1$. 
\end{remark}
\begin{remark}
	Recall Theorem \ref{thm:lb-CR} that states that for any causal policy $\pi$, $\textsc{cr}_{\pi}\ge c_1 P(c_2\hat{s})/P(c_3\hat{s})$, where $c_1,c_2,c_3>0$, $c_1>0.13$, $c_2-c_3\ge 0.14$, and $c_2/c_3\ge 1.07$. 
	Theorem \ref{thm:CR-bounds} shows that the dependence of $\textsc{cr}_{\pi^g}$ on the system parameters is similar to the lower bound on $\textsc{cr}_{\pi}$ in Theorem \ref{thm:lb-CR}. So, for any (convex) $P(s)$ and $\hat{s}$ (i.e., $W/D$), if there exists a causal policy $\pi$ with bounded competitive ratio $\textsc{cr}_{\pi}$, then $\textsc{cr}_{\pi^g}$ is also bounded.
\end{remark}

To prove Theorem \ref{thm:CR-bounds}, 
we need some structural results for an optimal offline policy $\pi^\star$, which we derive as follows.

 \section{Properties of an Optimal Offline Policy $\pi^\star$} \label{sec:opt-policy}
Consider an optimal offline policy $\pi^\star$. In this section, for simplicity, we only consider the packets that are transmitted by $\pi^\star$ in interval $[0,T]$, and index them 
as $1,2,3,...$ in ascending order of their generation times (i.e., $t_1<t_2<t_3<...$). 
Therefore, between the generation time of packets $i-1$ and $i$, many other packets might have been generated, however, they are not transmitted by $\pi^\star$. 
 \begin{lemma} \label{lemma:2W-per-D}
	If $\pi^\star$ chooses to transmit packet $i$, and $d_i\le T$ (where $d_i=t_i+D$), then in interval $[t_i,d_i)$, $\pi^\star$ transmits at least two packets completely (i.e., $2W$ bits). 
\end{lemma}
\begin{IEEEproof}
	Note that if $\pi^\star$ delivers packet $i$ at time $\tau_i$, then the deadline at time $\tau_i$ is $d(\tau_i)=t_i+D=d_i$. 
	Since $d_i\le T$, from Proposition \ref{prop:feasible}, it follows that $\pi^\star$ delivers packet $i+1$ in interval $[\tau_i,d_i)$ (i.e., $\tau_{i+1}\in[\tau_i,d_i)$). 
	Thus, in interval $[t_i,d_i)$, $\pi^\star$ transmits at least two complete packets $i$ and $i+1$ (i.e., $2W$ bits), as shown in Figure \ref{fig:typical-speed-profile} (for $i=2$). 
\end{IEEEproof} 
Thus, Lemma \ref{lemma:2W-per-D} implies that for $\pi^\star$, the intervals $[t_i,d_i)$ are special, and hence, we call them \emph{periods}, defined rigorously next. 
\begin{definition} \label{def:period}
	With respect to $\pi^\star$, the interval $\chi_i=[t_i,d_i)$ 
	is called \textbf{period} $i$. 
\end{definition}
By definition, a period starts at the generation time of a packet transmitted by $\pi^\star$, and in each period $i$, $\pi^\star$ transmits at least two packets (from Lemma \ref{lemma:2W-per-D}).  
Thus, consecutive periods overlap as shown in Figure \ref{fig:frames}. Hence, it is difficult to generalize Lemma \ref{lemma:2W-per-D} directly to the whole interval $[0,T]$. Therefore, we further define \emph{frames} (that are non-overlapping) with respect to $\pi^\star$ as follows.  
 \begin{definition} \label{def:frame}
	With respect to $\pi^\star$, the interval $I_i=[d_{i-1},d_i)$ (where $d_{i-1}=t_{i-1}+D$ and $d_i=t_i+D$) is called \textbf{frame} $i$.  
\end{definition}
\begin{remark} \label{remark:partition}
	As shown in Figure \ref{fig:frames}, consecutive frames partition the time-axis between $[0,T]$ (assuming the first frame $I_0$ starts at time $t=0$). Therefore, $\forall i\ne j$,  $I_i\cap I_j=\phi$, and there exists consecutive frames $0,1,...,m$, such that $T$ lies in frame $m$, and  $[0,T]\subseteq\cup_{i=1}^{m}I_i$. 
	So, the properties of $\pi^\star$ in a frame can be easily generalized to the entire interval $[0,T]$. 
\end{remark}

	In Definition \ref{def:period} and Definition \ref{def:frame}, note that periods and frames are defined with respect to the packets transmitted by an optimal offline policy $\pi^\star$. Also, note that frame $i$ (interval $[d_{i-1},d_i)$) is a subset of period $i$ (interval $[t_i,d_i)$), with $d_{i-1}>t_i$. Hence, length of a frame is always less than $d_i-t_i=D$.
\begin{figure} 
	\begin{subfigure} {0.5\textwidth}
		\begin{center}
			\begin{tikzpicture}[thick,scale=1, every node/.style={scale=1}]
				
				\draw[->] (0,0) node[below]{$t_2$} to (7,0) node[below]{$t$};
				\draw[->] (0,0) to (0,2.5) node[left]{$s(t)$};
				
				\draw (0,1) node[left]{$s_1$} to (0.4,1) to (0.4,0) node[below]{$\tau_1$};
				\filldraw[pattern=north east lines, pattern color=lightgray] (0.8,0) node[below]{$r_2$} to (0.8,2) to (2.3,2) to (2.3,0) node[below]{$\tau_2$};
				\node at (1.55,1) {$W$};
				\filldraw[pattern=north east lines, pattern color=lightgray] (3.4,0) node[below right]{$r_3$} to (3.4,1.5) to (5.4,1.5) to (5.4,0) node[below left]{$\tau_3$};
				\node at (4.4,0.75) {$W$};
				
				\draw (5.9,0) node[below]{$t_4,r_4$} to (5.9,0.5) to (6.6,0.5);
				
				\draw[dashed] (1.2,0) node[below]{$d_{0}$} to (1.2,2.1); 
				\draw[dashed] (2.75,0) node[below]{$d_1$} to (2.75,2.1);
				\draw[dashed] (3.2,0) node[below]{$t_3$} to (3.2,0.1);
				\draw[dashed] (6.6,0) node[below]{$d_2$} to (6.6,2.1);
				
				\draw[dashed] (0,2) node[left]{$s_2$} to (1,2);
				\draw[dashed] (0,1.5) node[left]{$s_3$} to (3.7,1.5);
				\draw[dashed] (0,0.5) node[left]{$s_4$} to (5.9,0.5);
				\draw[|<->|] (0,-0.65) -- (6.6,-0.65) node[rectangle,inner sep=-1pt,midway,fill=white]{$D$};
				\draw[|<->|] (1.2,2.25) -- (2.75,2.25) node[rectangle,inner sep=-1pt,midway,fill=white]{$I_{1}$};
				\draw[<->|] (2.75,2.25) -- (6.6,2.25) node[rectangle,inner sep=-1pt,midway,fill=white]{$I_{2}$};

			\end{tikzpicture}
			\caption{Typical speed profile in a period. Here, a packet generated at $t_i$ is scheduled for transmission at $r_i$, and delivered to the monitor at $\tau_i$.\vspace{1.5ex}} 
			\label{fig:typical-speed-profile} 
		\end{center} 
	\end{subfigure}
	\begin{subfigure}{0.5\textwidth}
	\begin{center}
		\begin{tikzpicture}[thick,scale=1, every node/.style={scale=1}]
		\draw[->] (0,0) node[above]{$0$} to (7.5,0) node[below]{time ($t$)};
		
        \draw (1.5,0) node[above]{$d_0$};
		\draw[very thick] (0.75,0) to (0.75,-0.5) to (2.75,-0.5) to (2.75,0);
		\draw[very thick] (1.75,0) to (1.75,0.5) node[above right]{$t_2$} to (3.75,0.5) node[above left]{$d_2$} to (3.75,0);
		\draw[very thick] (2.75,0) to (2.75,-0.5) to (4.75,-0.5) to (4.75,0);
		\draw[very thick] (4,0) to (4,0.5) node[above]{$t_4$} to (6,0.5) node[above]{$d_4$} to (6,0);
		\draw[very thick] (5,0) to (5,-0.5) to (7,-0.5); 
		
		\draw[dashed] (0,0) to (0,-1);
		\draw[dashed] (0.75,0) node[above]{$t_1$}; 
		\draw[dashed] (1.5,0) to (1.5,-1);
		\draw[dashed] (2.75,0.5) to (2.75,-1);
		\draw (2.75,0.5) node[above]{$d_1,t_3$};
		\draw[dashed] (3.75,0) to (3.75,-1);
		\draw[dashed] (4.75,0.5) node[above left]{$d_3$} to (4.75,-1);
		\draw[dashed] (5,0.5) node[above right]{$t_5$}; 
		\draw[dashed] (6,0) to (6,-1);
		\draw[dashed] (6.8,0) node[above]{$T$} to  (6.8,-1); 
		
		\draw[|<->|] (1.75,1.2) -- (3.75,1.2) node[rectangle,inner sep=-1pt,midway,fill=white]{$D$};
		\draw[|<->|] (4,1.2) -- (6,1.2) node[rectangle,inner sep=-1pt,midway,fill=white]{$D$};
		\draw[|<->|] (0.75,-1.5) -- (2.75,-1.5) node[rectangle,inner sep=-1pt,midway,fill=white]{$D$};
		\draw[<->|] (2.75,-1.5) -- (4.75,-1.5) node[rectangle,inner sep=-1pt,midway,fill=white]{$D$};
		
		\draw[|<->] (0,-1) -- 
		(1.5,-1) node[rectangle,inner sep=-1pt,midway,fill=white]{$I_{0}$};
		\draw[<->] (1.5,-1) -- (2.75,-1) node[rectangle,inner sep=-1pt,midway,fill=white]{$I_{1}$};
		\draw[<->] (2.75,-1) -- (3.75,-1) node[rectangle,inner sep=-1pt,midway,fill=white]{$I_{2}$};
		\draw[<->] (3.75,-1) -- (4.75,-1) node[rectangle,inner sep=-1pt,midway,fill=white]{$I_{3}$};
		\draw[<->] (4.75,-1) -- (6,-1) node[rectangle,inner sep=-1pt,midway,fill=white]{$I_{4}$};
		\draw[<->] (6,-1) -- (7,-1) node[rectangle,inner sep=-1pt,midway,fill=white]{$I_{5}$};
		
		\fill[pattern=north east lines, pattern color=lightgray] (0.75,0) to (1.5,0) to (1.5,-0.5) to (0.75,-0.5) to (0.75,0);
		\fill[pattern=north west lines, pattern color=lightgray] (1.75,0.5) to (2.75,0.5) to (2.75,-0.5) to (1.75,-0.5) to (1.75,0.5);
		\fill[pattern=north east lines, pattern color=lightgray] (2.75,0.5) to (3.75,0.5) to (3.75,-0.5) to (2.75,-0.5) to (2.75,0.5);
		\fill[pattern=north west lines, pattern color=lightgray] (4,0.5) to (4.75,0.5) to (4.75,-0.5) to (4,-0.5) to (4,0.5);
		\fill[pattern=north east lines, pattern color=lightgray] (5,0.5) to (6,0.5) to (6,-0.5) to (5,-0.5) to (5,0.5);
		
		\end{tikzpicture}
	\caption{Time-axis as union of disjoint intervals (frames).} 
	\label{fig:frames} 
		\end{center}
	\end{subfigure}

\caption{Periods and Frames.\vspace{-2ex}}
\end{figure}
Figure \ref{fig:typical-speed-profile} shows a typical relation between periods and frames, where at time $t_2$, the deadline is $d_0$, and packet $1$ is delivered to the monitor at time $\tau_1< d_0$. Thus the deadline is updated at time $\tau_1$ to $d_1$. Then, the interval $I_1=[d_0,d_1)$ is called frame $1$.  
Similarly, within interval $I_1$, packet $2$ (with generation time $t_2$) is delivered to the monitor at time $\tau_2 <d_1$. Therefore, at time $\tau_2$, the deadline gets updated to $d_2$, and the interval $I_2=[d_1,d_2)$ is called frame $2$. 
\begin{remark}
	Although Figure \ref{fig:typical-speed-profile} shows typical properties of frames and periods defined with respect to the packets transmitted by $\pi^\star$, the speed profile shown in Figure \ref{fig:typical-speed-profile} is not necessarily the speed chosen by $\pi^\star$, since that ($\pi^\star$) is unknown. 
	We show in Proposition \ref{prop:facts-min-energy-opt-frame} that within a frame, speed of $\pi^\star$ exhibits several structural properties. 
\end{remark} 
\begin{proposition} \label{prop:facts-min-energy-opt-frame}
	The optimal offline policy $\pi^\star$
		1) transmits the $W$ bits of a packet with constant speed,
		2) never interrupts transmission of any packet,
        3) delivers packet $i+1$ in frame $i$ ($\forall i$), and 
		4) never decreases the transmission speed within a frame.
\end{proposition}
\begin{IEEEproof} 
	See Appendix \ref{appendix:proof-prop-facts-min-energy-opt-frame}.
\end{IEEEproof}
\begin{remark}
	The third property in Proposition \ref{prop:facts-min-energy-opt-frame} follows due to the optimality of $\pi^\star$, and not from the definition of frames. To understand this, note that frame $i$ depends only on packet $i-1$ and $i$ transmitted by $\pi^\star$, and does not restrict packet $i+2$ from being transmitted in frame $i$, in addition to packet $i+1$.
\end{remark}

Next, Theorem \ref{thm:universal-lower-bound} below shows that for 
$\pi^\star$, 
the energy consumed is at least $P(3W/D)(T-D)$. 
\begin{theorem} \label{thm:universal-lower-bound}
	In interval $[0,T]$, the energy consumed by an optimal offline policy $\pi^\star$ 
	is at least $\max\{0,P(2W/D)(T-D)\}$ (irrespective of the packet generation sequence $\sigma$). 
\end{theorem}
\begin{IEEEproof}
	See Appendix \ref{appendix:universal-lower-bound}.
\end{IEEEproof}
\begin{remark}
	Note that the lower bound on the energy consumption 
	provided in Theorem \ref{thm:universal-lower-bound} is independent of the actual packet generation times. Hence, we call it \emph{Universal Lower Bound} (ULB). Since ULB is agnostic of actual packet generation times, it is too weak for proving Theorem \ref{thm:CR-bounds}. Nonetheless, 
	it helps in understanding the dependence of energy consumption of an optimal offline policy $\pi^\star$ on critical parameters like packet size $W$ and peak AoI $D$. 
\end{remark}

\section{Proof of Theorem \ref{thm:CR-bounds}} \label{sec:proof-thm-CR-bounds}
We prove Theorem \ref{thm:CR-bounds} in two steps. In step 1, we show that $\textsc{cr}_{\pi^g} \le 2P(3\hat{s})/P(\hat{s})+1$, while in step 2, we show that $1.5P(3\hat{s})/P(1.5\hat{s})\le \textsc{cr}_{\pi^g}$.

\subsection*{Step 1: Upper Bound on $\textsc{cr}_{\pi^g}$} 
Consider an arbitrary sequence of packet generation times $\sigma$.
From Remark \ref{remark:partition}, we know that the time axis can be partitioned into frames defined with respect to $\pi^\star$ (Definition \ref{def:frame}).  
Therefore, consider the consecutive frames $0,1,...,m$ such that the total time horizon interval $[0,T]$ is a subset of the union of frames $0$ to $m$, and time horizon $T$ lies in frame $m$ (as shown in Figure \ref{fig:frames} for $m=5$). Note that if $m=0$ (i.e., $T$ is less than the initial deadline $d(0)$), then neither $\pi^g$, nor $\pi^\star$ transmits any packet because the deadline constraint \eqref{eq:deadline-constraint} is trivially satisfied in the interval $[0,T]$. Hence, we only consider the case where $m\ge 1$.

Since length of a frame is always less than $D$ (the length of a period that is equal to the peak AoI constraint), the time horizon $T<(m+1)D$. Therefore, in interval $[0,T]$, if $\pi^g$ transmits $x^g \ge 0$ number of packets with speed $3W/D$ consuming $E^g_x$ units of energy, 
then $E^g_x<P(3W/D)(m+1)D$ (product of power consumption and upper bound on the length of time interval $[0,T]$). 
So, if $y^g \ge 0$ denotes the number of packets that $\pi^g$ transmits with speed greater than $3W/D$ (recall that $\pi^g$ transmits an entire packet at a constant speed), and $E^g_y$ denotes the total energy consumed by $\pi^g$ in transmitting these $y^g$ packets,  
then the total energy consumed by $\pi^g$ in interval $[0,T]$ is \footnote{Note that at any time $t$, $\pi^g$ either transmits packets at speed greater than or equal to $3W/D$, or remains idle (speed is $0$). As per Assumption \ref{assume:P0}, when $\pi^g$ is idle, power consumption is $0$.} 
\begin{align} \label{eq:energy-pig-ub}
	E_{\pi^g}=E^g_x+E^g_y\le (m+1)P(3W/D)D+E^g_y.
\end{align}
\begin{remark} \label{remark:per-pkt-energy-greater-speed}
	Since $\pi^g$ transmits each of the $y^g$ packets at a constant speed greater than $3W/D$, the energy consumed by $\pi^g$ in transmitting the $y^g$ number of packets is $E^g_y>y^gP(3W/D)D/3$. 
\end{remark}

From Proposition \ref{prop:facts-min-energy-opt-frame} (Property 3), it follows that $\pi^\star$ delivers exactly one packet in each frame $i=0,1,2,...,m-1$ ($\pi^\star$ does not transmit any packet in frame $m$ because by definition, $T$ lies in frame $m$, and hence, at the start of frame $m$, the deadline (end of frame $m$) is already greater than $T$). Thus, $\pi^\star$ transmits $m$ packets in interval $[0,T]$. 
Also, the length of each frame is less than $D$ (length of a period). Hence, the energy consumed by $\pi^\star$ in transmitting each of these $m$ packets is at least $P(W/D)D$. 
Next, we show in Lemma \ref{lemma:opt-consumes-E^g_y} that out of these $m$ packets that $\pi^\star$ transmits completely, there exists a subset $\cZ$ consisting of $y^g$ number of packets, such that $\pi^\star$ consumes at least $E^g_y$ units of energy in transmitting the packets in $\cZ$ (where $y^g$ and $E_y^g$ are defined as in Remark \ref{remark:per-pkt-energy-greater-speed}).
\begin{lemma} \label{lemma:opt-consumes-E^g_y}
	There exists a subset $\cZ$ consisting of $y^g$ number of packets such that $\pi^\star$ transmits all the packets in $\cZ$, and consumes at least $E^g_y$ units of energy while transmitting the packets in $\cZ$.
\end{lemma}
\begin{IEEEproof}
	Note that for $y^g=0$, the claim is trivially satisfied. So, for the rest of the proof, we assume $y^g\ge 1$. 
	Recall that $\pi^g$ transmits $y^g$ number of packets at speed greater than $3W/D$, consuming $E^g_y$ units of energy. Without loss of generality, let the packets be indexed as $1,2,...,y^g$. Also, let $\pi^g$ transmits a packet $j\in\{1,2,...,y^g$\} over the time interval $U_j$, and consumes energy $e_j$ (in transmitting packet $j$). Since $\pi^g$ transmits only one packet at a time, $U_i\cap U_j=\phi$ for $i\ne j$. Therefore, to prove Lemma \ref{lemma:opt-consumes-E^g_y}, it is sufficient to show that in interval $U_j$ (for $j\in\{1,2,...,y^g\}$), $\pi^\star$ transmits at least one packet completely (entire $W$ bits), consuming at least $e_j$ units of energy. This follows from Lemma \ref{lemma:partitionAB} (in Appendix \ref{appendix:lemma-partitionAB}), where we show that for each packet $j$ that $\pi^g$ transmits with speed greater than $3W/D$, $\pi^\star$ transmits at least one packet $\hat{j}$ (where packet $j$ and $\hat{j}$ may be same) completely during the time interval when $\pi^g$ transmits packet $j$, at a constant speed at least equal to the constant speed with which $\pi^g$ transmits packet $j$.
\end{IEEEproof}
\begin{remark}
	$\pi^\star$ transmits a total of $m$ packets in interval $[0,T]$ (exactly one packet in each of the frames $0,1,2,...,m-1$). Also, from Lemma \ref{lemma:opt-consumes-E^g_y}, it follows that $\pi^\star$ transmits at least $y^g$ number of packets. Therefore, $y^g\le m$. 
\end{remark}
Hence, the total energy consumed by $\pi^\star$ in interval $[0,T]$ is 
\begin{align} \label{eq:energy-pistar-lb}
	E_{\pi^\star}\ge (m-y^g)P(W/D)D+E^g_y. 
\end{align} 

From \eqref{eq:energy-pig-ub} and \eqref{eq:energy-pistar-lb}, we obtain an upper bound on the competitive ratio \eqref{eq:CR-definition} for $\pi^g$ as follows. 
\begin{align}
	\textsc{cr}_{\pi^g}&\le \frac{(m+1)P(3W/D)D+E^g_y}{(m-y^g)P(W/D)D+E^g_y}, \nonumber \\ &\le \frac{(m+1)P(3W/D)D}{(m-y^g)P(W/D)D+E^g_y}+1, \nonumber \\
	&\stackrel{(a)}{\le} \frac{(m+1)P(3W/D)D}{(m-y^g)P(W/D)D+y^gP(3W/D)D/3}+1, \nonumber \\ \label{eq:RHSofCRUB}
	&\stackrel{(b)}{\le} \frac{(m+1)P(3W/D)}{mP(W/D)}+1,  \\
	&\stackrel{(c)}{\le} \frac{2P(3W/D)}{P(W/D)}+1,
\end{align}
where in $(a)$, we have used the fact that $E^g_y>y^gP(3W/D)D/3$ (Remark \ref{remark:per-pkt-energy-greater-speed}), 
$(b)$ follows because $P(3W/D)D/3 \ge P(W/D)D$ (transmitting $W$ bits at speed $3W/D$ consumes more energy than transmitting $W$ bits at speed $W/D$), and we get $(c)$ by maximizing the R.H.S. of \eqref{eq:RHSofCRUB} with respect to $m\ge 1$. 
With $\hat{s}=W/D$, we get the result.

\subsection*{Step 2: Lower Bound on $\textsc{cr}_{\pi^g}$}
	Let $\delta\to 0^+$, and for a fixed $D$, consider the following problem instance, where
	$T=4D/3+\delta/2$, initial AoI $\Delta(0)=0$, and 
	three packets are generated at time $t=0$, $t=D/3$ and $t=D/3+\delta$, respectively. Optimal offline policy $\pi^\star$ transmits only the third packet, with speed $W/(2D/3-\delta)$ over the interval $[D/3+\delta,D)$. On the other hand, 
	$\pi^g$ transmits all three packets with speed $3W/D$, over the intervals $[0,D/3)$, $[D/3,2D/3)$ and $[2D/3,D)$, respectively. Therefore, 
	$\textsc{cr}_{\pi^g}\ge 
	3P(3W/D)(D/3)/(P(W/(2D/3-\delta))(2D/3-\delta))\to 1.5P(3\hat{s})/P(1.5\hat{s})$, where $\hat{s}=W/D$.

Thus far, we analyzed the base model considered in Section \ref{sec:SystemModel}. Next, in Sections \ref{sec:power-constraint} and \ref{sec:arbitrary-sizes}, we consider two relevant generalizations of the base model, and derive some structural results.

\section{Peak Transmit Power Constraint} \label{sec:power-constraint}
In this section, we consider the setting where in addition to the peak AoI constraint \eqref{eq:constraint} (i.e., the deadline constraint \eqref{eq:deadline-constraint}), a policy must also satisfy a peak transmit power constraint.

To model the peak transmit power constraint, at any time, let the maximum power with which the node can transmit be $\beta$, i.e., $P(s(t))\le \beta$, $\forall t\in[0,T]$. 
Since $P(\cdot)$ is an increasing function, $P^{-1}(\beta)$ is well defined, and the peak power constraint $P(s(t))\le \beta$ is equivalent to the maximum speed constraint: $s(t)\le s^{\max}$, $\forall t\in[0,T]$, where $s^{\max}=P^{-1}(\beta)$.
With this additional constraint, the optimization problem \eqref{eq:objective}--\eqref{eq:deadline-constraint} becomes:
\begin{subequations}
	\begin{align} \label{eq:p-objective}
		&\underset{\pi\in\Pi}{\min}\ \ E_{\pi}(\sigma) = \int_{t=0}^{T}P(s(t))dt \\
		\label{eq:p-deadline-constraint}
		&\text{s.t.}\ \  d(t)> t, \ \ \forall t\in [0,T], \\
		\label{eq:p-speed-constraint}
		&\hspace{2.5ex}\ \  s(t)\le s^{\max}, \ \ \forall t\in [0,T]. 
	\end{align}
\end{subequations}

Note that in previous sections, where we did not have any power constraint, under the assumption that the inter-generation time of packets is less than $D$ (Remark \ref{remark:X<D}), the deadline constraint \eqref{eq:p-deadline-constraint} was always feasible (the problem had a feasible solution). However, with peak power/speed constraint \eqref{eq:p-speed-constraint}, the sufficient conditions for feasibility of the deadline constraint \eqref{eq:p-deadline-constraint} are non-trivial. 
We illustrate this in Appendix \ref{appendix:example_speed_constraint} 
where using an example we show that under the speed constraint \eqref{eq:p-speed-constraint}, it may be possible to satisfy the deadline constraint \eqref{eq:p-deadline-constraint} for a sequence of packets with some constant inter-generation time $a$, but not for a sequence of packets with constant inter-generation time $a'<a$.  
This is against the general intuition that smaller inter-generation times are better for AoI minimization. 
Thus, determining the feasibility of optimization problem \eqref{eq:p-objective}--\eqref{eq:p-speed-constraint} based on the packet inter-generation times is a non-trivial task. 
Therefore, we define a set of  packet generation sequences $\Sigma(s^{\max})$ as follows (Definition \ref{def:Sigma}), and analyze the optimization problem \eqref{eq:p-objective}--\eqref{eq:p-speed-constraint} for 
sequences $\sigma\in\Sigma(s^{\max})$, for different values of $s^{\max}$. 
\begin{definition} \label{def:Sigma}
	For any $s^{\max}$, let
	$\Sigma(s^{\max})$ denote the set of all packet generation sequences $\sigma$ for which there exists some offline policy $\pi(\sigma)$ that can satisfy the constraints \eqref{eq:p-deadline-constraint} and \eqref{eq:p-speed-constraint} simultaneously. 
	By definition, for any $\sigma\not\in\Sigma(s^{\max})$, constraints \eqref{eq:p-deadline-constraint} and \eqref{eq:p-speed-constraint} can never be satisfied simultaneously (i.e. the optimization problem \eqref{eq:p-objective}--\eqref{eq:p-speed-constraint} has no feasible solution).
\end{definition}
%

Next, we analyze problem \eqref{eq:p-objective}--\eqref{eq:p-speed-constraint} in three parts based on the value of $s^{\max}$.

\subsection{$s^{\max}\ge 3W/D$}
In this case, we consider the policy $\pi^g$ defined earlier for the case when there is no upper bound on the transmit power at any time with no modification. 
Using the structure of the packet generation sequence set $\Sigma(s^{\max})$,
we show next that for any 
$\sigma\in\Sigma(s^{\max})$, $\pi^g$ never needs to choose a speed greater than $s^{\max}$. Because of this automatic satisfaction of the speed constraint (peak transmit power constraint) by $\pi^g$, we will also get that $\pi^g$ satisfies the peak AoI constraint 
and has competitive ratio (defined with respect to the set 
$\Sigma(s^{\max})$) equal to
$\textsc{cr}_{\pi^g}\le 2P(3\hat{s})/P(\hat{s})+1$.

\begin{lemma} \label{lemma:smax>3W/D}
	For any $s^{\max}\ge 3W/D$ and 
	sequence $\sigma\in\Sigma(s^{\max})$, the greedy policy $\pi^g$ (Algorithm \ref{algo:greedy}) satisfies both the constraints \eqref{eq:p-deadline-constraint} and \eqref{eq:p-speed-constraint} simultaneously. Also, over the set of packet generation sequences $\Sigma(s^{\max})$, $\pi^g$ has a competitive ratio of $\textsc{cr}_{\pi^g}\le\frac{2P(3\hat{s})}{P(\hat{s})}+1$ (where $\hat{s}=W/D$). 
\end{lemma}
\begin{remark}
	Note that for policy $\pi^g$, the competitive ratio upper bound in Lemma \ref{lemma:smax>3W/D} is same as that in Theorem \ref{thm:CR-bounds} (i.e. when there is no speed constraint \eqref{eq:p-speed-constraint}). 
\end{remark}
\begin{IEEEproof}[Proof of Lemma \ref{lemma:smax>3W/D}]
	Since $\pi^g$ always satisfies the deadline constraint \eqref{eq:p-deadline-constraint}, we have that $\pi^g$ satisfies both the deadline constraint \eqref{eq:p-deadline-constraint} and speed constraint \eqref{eq:p-speed-constraint} simultaneously if the speed \eqref{eq:speed} under $\pi^g$ never exceeds $s^{\max}$. Therefore, we next show that when $s^{\max}\ge 3W/D$, for packet generation sequences $\sigma\in\Sigma(s^{\max})$, the speed under $\pi^g$ never exceeds $s^{\max}$.

Consider any packet generation sequence $\sigma\in\Sigma(s^{\max})$, and a causal policy $\pi(\sigma)$ that satisfies the constraints \eqref{eq:p-deadline-constraint} and \eqref{eq:p-speed-constraint} simultaneously for $\sigma$.
Recall the proof of Lemma \ref{lemma:partitionAB} (from Appendix \ref{appendix:lemma-partitionAB}), where we have shown that corresponding to each packet that the policy $\pi^g$ transmits at speed $s^g>3W/D$, there exists a packet that any policy $\pi$ that satisfies the deadline constraint \eqref{eq:p-deadline-constraint}, must transmit at average speed (maximum speed) at least equal to $s^g$. Therefore, if $\pi^g$ does not satisfy the speed constraint \eqref{eq:p-speed-constraint} (i.e., at some time, $\pi^g$ transmits at speed $s^g>s^{\max}\ge 3W/D$), then no policy can satisfy both \eqref{eq:p-deadline-constraint} and \eqref{eq:p-speed-constraint} simultaneously. But by definition, for $\sigma\in\Sigma(s^{\max})$, $\pi(\sigma)$
satisfies both \eqref{eq:p-deadline-constraint} and \eqref{eq:p-speed-constraint} simultaneously.
Hence, $\pi^g$ must also satisfy \eqref{eq:p-speed-constraint} (never transmit at speed greater than $s^{\max}$). 

Further, note that the optimization problem \eqref{eq:p-objective}--\eqref{eq:p-speed-constraint} differs from the optimization problem \eqref{eq:objective}--\eqref{eq:deadline-constraint} only in the constraint \eqref{eq:p-speed-constraint}. Since an additional constraint cannot decrease the energy consumption for an optimal offline policy $\pi^\star$, and the policy $\pi^g$ 
is unchanged compared to the case when there is no speed constraint, the competitive ratio upper bound \eqref{eq:CR-bounds} derived for $\pi^g$ in Theorem \ref{thm:CR-bounds} is still valid.
\end{IEEEproof} 


\subsection{$s^{\max}\in[2W/D,3W/D)$}
\begin{lemma} \label{lemma:no-feasible-policy}
	When $s^{\max}\in[2W/D,3W/D)$, no causal policy $\pi$ can satisfy the deadline constraint \eqref{eq:p-deadline-constraint} and the speed constraint \eqref{eq:p-speed-constraint}, simultaneously for all 
	input $\sigma\in\Sigma(s^{\max})$.
\end{lemma}
\begin{IEEEproof}[Proof Sketch]
	For any fixed	$s^{\max}\in[2W/D,3W/D)$, we consider a pair of inputs $\sigma_1,\sigma_2\in\Sigma(s^{\max})$ with a common prefix, i.e., in time interval $[0,\tau]\subseteq [0,T]$, packet generation times for $\sigma_1$ and $\sigma_2$ are the same. We will show that for any causal policy $\pi$ that satisfies the maximum speed constraint \eqref{eq:p-speed-constraint}, 
	if the deadline constraint \eqref{eq:p-deadline-constraint} is satisfied in interval $[0,\tau]$ (for $\sigma_1$, $\sigma_2$), then in interval $(\tau,T]$, the deadline constraint \eqref{eq:p-deadline-constraint} will be violated by $\pi$ for at least one of $\sigma_1$ and $\sigma_2$. 
	For details, see Appendix \ref{appendix:proof-lemma-no-feasible-policy}.
\end{IEEEproof}

Lemma \ref{lemma:no-feasible-policy} shows that when $s^{\max}\in[2W/D,3W/D)$, for any causal policy $\pi$, there exists some packet generation sequence $\sigma\in\Sigma(s^{\max})$ for which $\pi$ violates the constraints \eqref{eq:p-deadline-constraint} or \eqref{eq:p-speed-constraint}. Hence, when $s^{\max}\in[2W/D,3W/D)$, the competitive ratio of all causal policies is unbounded. 

\subsection{$s^{\max}< 2W/D$} \label{sec:Smax<2W/D} 
\begin{remark} \label{remark:d(0)>T}
	Note that if the initial deadline $d(0)>T$, then an optimal causal/offline policy should not transmit any packet (since constraints \eqref{eq:p-deadline-constraint} and \eqref{eq:p-speed-constraint} are automatically satisfied, and energy consumption is $0$). Hence, in this subsection, we consider the only non-trivial setting where $d(0)\le T$, and any policy that satisfies the deadline constraint \eqref{eq:p-deadline-constraint} must transmit at least one packet completely until time $d(0)$.
\end{remark}

\begin{lemma} \label{lemma:g_i>T-D}
	 Consider an offline policy $\pi$ that satisfies constraints \eqref{eq:p-deadline-constraint} and \eqref{eq:p-speed-constraint}. For $s^{\max}<2W/D$ and packet generation sequence $\sigma\in\Sigma(s^{\max})$, 
	 $\pi$ completely transmits a packet $i$ only if its generation time $g_i>T-D$. 
\end{lemma} 
\begin{IEEEproof}
	Recall the proof of Lemma \ref{lemma:2W-per-D}, where we have shown that if a causal/offline policy $\pi$ completely transmits a packet generated at time $g_i$, and $d_i=g_i+D\le T$, then to satisfy the deadline constraint \eqref{eq:p-deadline-constraint}, $\pi$ must transmit another packet $j$ with generation time $g_j>g_i$, completely before $d_i=g_i+D$. In other words, in interval $[g_i,g_i+D)$, $\pi$ must transmit $2W$ bits ($W$ bits each for packet $i$ and $j$). However, any policy that satisfies the speed constraint \eqref{eq:p-speed-constraint} (transmits at speed less than $s^{\max}<2W/D$) cannot transmit $2W$ bits in the interval $[g_i,g_i+D)$ of length $D$. Therefore, any offline policy that satisfies both the deadline constraint \eqref{eq:p-deadline-constraint} and speed constraint \eqref{eq:p-speed-constraint} simultaneously, must only transmit a packet $i$ if $d_i=g_i+D<T$, i.e. the packet $i$'s generation time $g_i<T-D$.
\end{IEEEproof}

%
Note that if policy $\pi$ completes transmitting a packet $i$ (with generation time $g_i>T-D$) at time $\tau_i>g_i$, then at any time $t\in[\tau_i,T]$, the deadline $d(t)=g_i+D>T\ge t$. Hence, once a packet $i$ ($g_i>T-D$) gets completely transmitted, an optimal offline policy $\pi^\star$ (that consumes minimum energy and satisfies constraints \eqref{eq:p-deadline-constraint} and \eqref{eq:p-speed-constraint}), should not transmit any other packet. This fact together with Remark \ref{remark:d(0)>T} and Lemma \ref{lemma:g_i>T-D} implies that in interval $[0,T]$, $\pi^\star$ completely transmits exactly one packet $i^\star$. Also, the generation time of $i^\star$ is $g_{i^\star}>T-D$, and the transmission completion time is $\tau_{i^\star}<d(0)$. 
Thus, using the property of increasing and convex power function $P(\cdot)$ (Lemma \ref{lemma:constant-speed-better}), we 
get the following result.
\begin{lemma} \label{lemma:opt-offline}
	For $s^{\max}<2W/D$ and packet generation sequences $\sigma\in\Sigma(s^{\max})$, an optimal offline policy $\pi^\star$ completely transmits exactly one packet $i^{\star}=\arg\min_{i}\{g_i|g_i>T-D\}$ with constant speed $W/(d(0)-g_{i^\star})$ over the interval $[g_{i^\star},d(0))$. 
\end{lemma}  


Lemma \ref{lemma:opt-offline} characterizes an optimal offline policy $\pi^\star$ for all packet generation sequences $\sigma\in\Sigma(s^{\max})$. However, the set $\Sigma(s^{\max})$ is non-empty (the deadline constraint \eqref{eq:p-deadline-constraint} is satisfied) only if the time horizon $T<3D/2$, as shown next.
\begin{corollary} \label{cor:T<3D/2}
	When $s^{\max}<2W/D$, the deadline constraint \eqref{eq:p-deadline-constraint} can be satisfied only if $T<3D/2$ (i.e. $D>2T/3$). 
\end{corollary}
\begin{IEEEproof} 
	Since $d(0)\le T$ (from Remark \ref{remark:d(0)>T}), to satisfy the deadline constraint \eqref{eq:p-deadline-constraint}, $\pi^\star$ must transmit a fresh packet completely before $d(0)$. Also, as per Lemma \ref{lemma:opt-offline}, $\pi^\star$ completely transmits only one packet $i^\star$. Hence, we get that under $\pi^\star$, the transmission of packet $i^\star$ completes before the initial deadline $d(0)\le D$. 
	
	Now, since $s^{\max}<2W/D$ (time required to completely transmit a packet is greater than $D/2$), and the transmission of packet $i^\star$ completes before $d(0)$, we get that the generation time of packet $i^\star$ is $g_{i^\star}< d(0)-D/2\le D-D/2=D/2$. Also, by definition, $g_{i^\star}\ge T-D$ (Lemma \ref{lemma:opt-offline}). Hence, we get $T-D\le g_{i^\star}<D/2$, which implies $T<3D/2$.
\end{IEEEproof}

Next, we show that the policy $\pi^\star$ (in Lemma \ref{lemma:opt-offline}) can in fact be implemented causally, and hence is an optimal causal policy. 
\begin{lemma} \label{lemma:OPT-smax}
	Optimal offline policy $\pi^\star$ characterized in Lemma \ref{lemma:opt-offline} can be implemented causally. 
\end{lemma}
\begin{IEEEproof}
	Consider the following causal threshold policy $\pi_{th}$:
	idle until a packet $i$ is generated with generation time $g_i>T-D$, and then begin to transmit packet $i$ (immediately) at constant speed $W/(d(0)-g_i)$ until completion, and then again idle until the time horizon $T$. 
	By definition, 
	identical to policy
	$\pi^\star$ defined in Lemma \ref{lemma:opt-offline}, $\pi_{th}$ transmits exactly one packet $i^{\star}=\arg\min_{i}\{g_i|g_i>T-D\}$ with constant speed $W/(d(0)-g_{i^\star})$ over the interval $[g_{i^\star},d(0))$. Therefore, the causal policy $\pi_{th}$ and $\pi^\star$ are identical, and they incur equal cost. 
\end{IEEEproof}

To conclude, in this section, we analyzed problem \eqref{eq:objective-init}--\eqref{eq:constraint} under the peak power constraint by splitting it into $3$ cases. We showed that
i) when $s^{\max}\ge 3W/D$, the competitive ratio of greedy policy $\pi^g$ (Algorithm \ref{algo:greedy}) is identical to the case when there is no power constraint, ii) when $s^{\max}\in[2W/D,3W/D)$, the competitive ratio of all causal policies is unbounded, and iii) when $s^{\max}<2W/D$, a threshold policy (Lemma \ref{lemma:OPT-smax}) is optimal with competitive ratio $1$.

\section{Arbitrary Packet Sizes} \label{sec:arbitrary-sizes}
In Section \ref{sec:SystemModel}, we assumed that each packet generated at the node is of size $W$ bits, where $W$ is a constant. However, this need not be true in practice. So, if the system model considered in Section \ref{sec:SystemModel} (without power/speed constraint) is relaxed to allow arriving packets to be of arbitrary sizes, then we have the following results.
\begin{theorem} \label{thm:arbitrary-packet-size}
	Among all the packets generated in the interval $[0,T]$, let the size of the smallest packet be $w$ bits, and the size of the largest packet be $W$ bits. Define $\zeta=W/w$. 
	\begin{enumerate}
		\item The competitive ratio of the greedy policy $\pi^g$ (Algorithm \ref{algo:greedy}) that uses $W$ for defining the speed is $\textsc{cr}_{\pi^g}\le 2P(3\zeta w/D)/P(w/D)+1$.
		\item When $\zeta\to\infty$, the competitive ratio of any causal policy is unbounded.
	\end{enumerate}
\end{theorem}
\begin{IEEEproof}
	See Appendix \ref{appendix:proof-thm:arbitrary-packet-size}.
\end{IEEEproof}

The second result of Theorem \ref{thm:arbitrary-packet-size} implies that if the packet sizes can vary arbitrarily, then the competitive ratio of any causal policy is unbounded. This is primarily because unlike causal policies, an optimal offline policy $\pi^\star$ knows the generation time of all the packets in advance. 
Therefore, $\pi^\star$ can avoid transmitting large size packets, while satisfying the peak AoI constraint \eqref{eq:constraint}. 
If in addition to packets having arbitrary sizes, it is required that each generated packet has to be transmitted, then the optimal offline policy $\pi^\star$ no longer has this advantage (i.e., $\pi^\star$ cannot avoid transmitting large sized packets). In fact, if the packets are required to be transmitted on First-Come-First-Serve (FCFS) basis\footnote{In context of AoI, transmitting packets on FCFS basis is meaningful, because receiving an old packet out of order, in general, is not useful.}, then a causal transmission policy with bounded competitive ratio can be obtained for particular power functions $P(s)$ as follows. 
 \begin{theorem} \label{thm:LOYDS-CR}
 	In the optimization problem \eqref{eq:objective} with deadline constraint \eqref{eq:deadline-constraint} and arbitrary packet sizes, if an additional FCFS constraint is imposed and all packets have to be delivered to the monitor, then for power function $P(s)=s^\alpha$ $(\alpha>1)$, the competitive ratio of Algorithm \ref{algo:LOYDS} has competitive ratio at most $\alpha^\alpha$.
 \end{theorem}
\begin{IEEEproof}
	To prove Theorem \ref{thm:LOYDS-CR}, we model the optimization problem \eqref{eq:objective} with deadline constraint \eqref{eq:deadline-constraint} and FCFS constraint, as an equivalent job scheduling problem, and use the existing results \cite{bansal2007speed} for job scheduling problems to conclude the proof.	
	
	When all the packets generated in interval $[0,T]$ are to be transmitted on FCFS basis, at the instant the node begins to transmit a packet $i$ (generated at time $t_i<T$), the latest packet delivered to the monitor is packet $i-1$. Hence, the deadline for packet $i$ is always $d_{i-1}=t_{i-1}+D$. 
	So, each packet $i$ can be considered as a job that needs to be processed before the deadline $d_{i-1}$, such that the overall energy consumption is minimized. Note that for any two distinct packets $i$ and $j$ with generation times $t_i$ and $t_j$, respectively, if $t_i<t_j$, then $t_{i-1}<t_{j-1}$ (packets are indexed in increasing order of their generation time). Therefore, when $t_i<t_j$, the deadline $d_i<d_j$, i.e., the deadline for the packet that is generated first, is earlier in time. Hence, Algorithm \ref{algo:LOYDS}, which at any time $t$, transmits the packet with earliest deadline, always transmits packets in FCFS order. 
	
	This equivalent job scheduling problem is a special case of the problem considered in \cite{yao1995scheduling}, and when the power function is $P(s)=s^\alpha$ ($\alpha>1$), \cite{bansal2007speed} showed that for the problem in \cite{yao1995scheduling}, Algorithm \ref{algo:LOYDS} has competitive ratio at most $\alpha^\alpha$ (irrespective of packet sizes). 
\end{IEEEproof}

\begin{algorithm}
	\caption{Modified Greedy Policy} 
	\label{algo:LOYDS}
	\begin{algorithmic}
		\STATE At any time $t$, transmit the available (undelivered) packet with earliest deadline, at speed
		$$s(t)=\max_{i\ge 1}\frac{w(t,t_i)}{(d_{i-1})-t},$$
		where $t_i\le t$ denotes the generation time of packet $i$, 
		$d_{i-1}=t_{i-1}+D$, and $w(t,t_i)$ denotes the sum of the number of undelivered bits of all the packets at time $t$, generated in interval $[0,t_i]$. 
	\end{algorithmic}
\end{algorithm} 

\begin{remark}
	Essentially, at any time $t$, Algorithm \ref{algo:LOYDS} assumes that no packet will be generated in the future, and transmits the available packets with speed such that the deadline constraint \eqref{eq:deadline-constraint} for the available packets are satisfied, while consuming minimum energy.
\end{remark}

\section{Numerical Results}
\begin{figure} 
	\centerline{
		\includegraphics[width=0.5\linewidth]{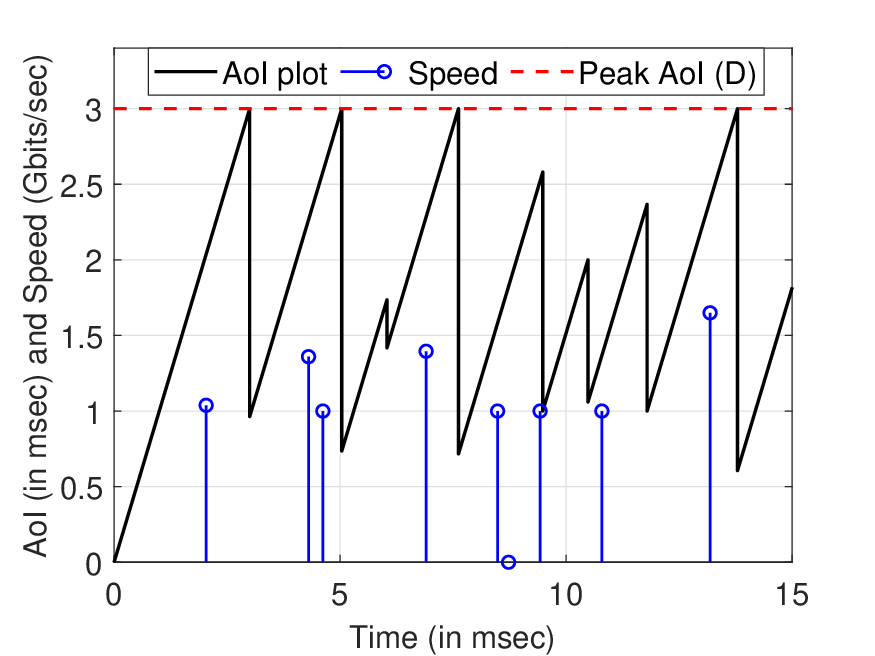}} 
	\caption{\small{AoI plot and transmission speed of packets.}}
	\label{fig:AoI_Speed}
\end{figure}

Figure \ref{fig:AoI_Speed} shows the AoI plot for $\pi^g$ when size of packets is $W=1$ Mbit, peak AoI is $D=3$ msec, and inter-generation time $X$ of packets follow uniform distribution with values in interval $(0,2.5)$.\footnote{\label{fn:feasible-problem}For the optimization problem \eqref{eq:objective} to be feasible, maximum inter-generation time must be less than $D$} Also, the stem plot in Figure \ref{fig:AoI_Speed} shows the generation time of packets, and the speed with which they were transmitted by $\pi^g$. The transmission speed of a packet is 0 if it is not transmitted, otherwise the speed is at least $3W/D=1$ Gbit/sec. Note that the speed is large (greater than 1) only if inter-generation time of packets is large (greater than $2D/3$). 

\begin{figure} 
	\centerline{
		\includegraphics[width=0.5\linewidth]{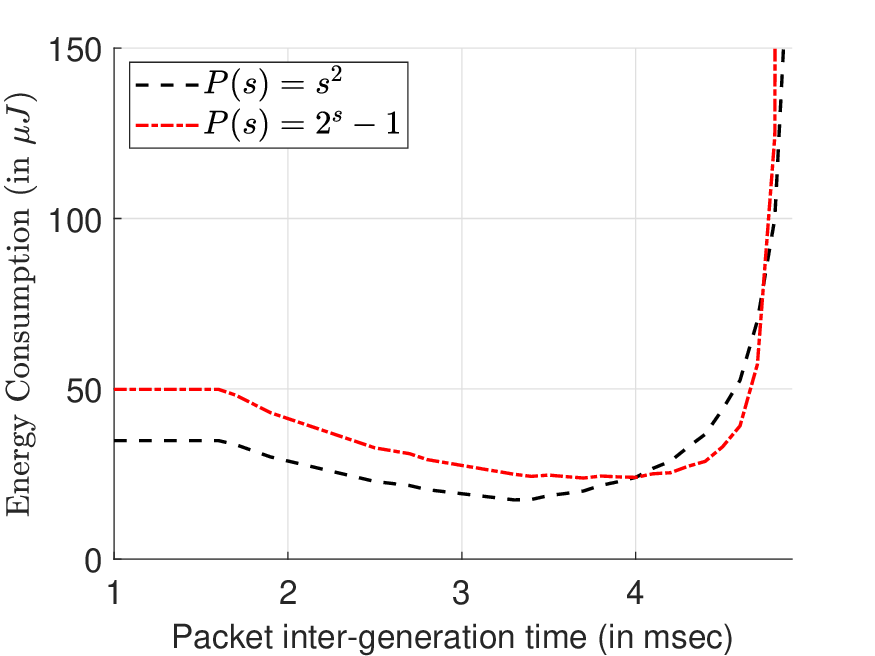}} 
	\caption{\small{Energy consumption as a function of inter-generation time.}}
	\label{fig:E-X-multiFn}
\end{figure}

To further understand the effect of inter-generation time of packets on energy consumption, Figure \ref{fig:E-X-multiFn} plots the energy consumed by $\pi^g$ as a function of inter-generation time $X$ (where $X$ is deterministic, packets are of size $W=1$ Mbit, and peak AoI and time horizon are $D=5$ msec and $T=100$ msec respectively). 
When $X\le D/3\approx 1.7$ msec, $\pi^g$ always has a fresh packet to transmit, and hence, remains busy throughout the interval $[0,T]$ transmitting packets with speed $3W/D=3/5$ Gbits/sec. So, as long as $X\le D/3$, energy consumption remains constant. When $X\in(D/3,2D/3]$, $\pi^g$ transmits fewer packets, but with same speed $3W/D$, and hence, consumes lesser energy. However, when $X>2D/3$, $\pi^g$ transmits fewer packets, but at speed larger than $3W/D$. So, energy consumption starts to increase with increase in $X$, and becomes unbounded at values of $X$ close to $D$.

\begin{remark}
	In all the simulations, when the speed $s$ is in Gbits/sec, we consider the power consumption $P(s)=s^2$ or $P(s)=2^s-1$ to be in milli-Watts.
\end{remark}

\begin{figure} 
	\centerline{
		\includegraphics[width=0.5\linewidth]{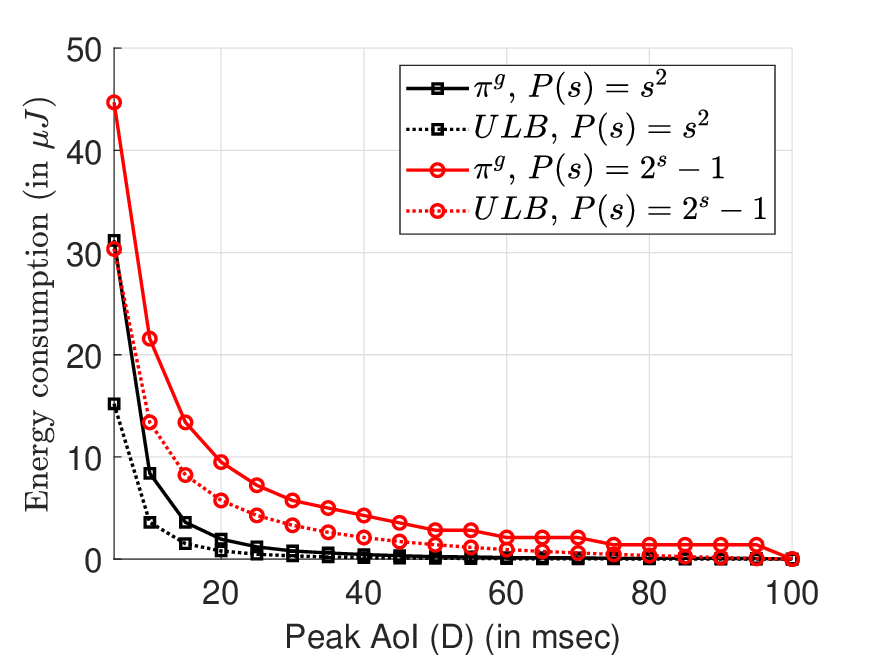}} 
	\caption{\small{Energy consumption as a function of peak AoI ($D$), and the universal lower bound (ULB) (Theorem \ref{thm:universal-lower-bound}).}} 
\label{fig:E_vs_D_multiFn}
\end{figure}

Next, we analyze the effect of $D$ (maximum allowed AoI) on the energy consumption by $\pi^g$. Note that if $D$ is large, $\pi^g$ gets more time to transmit individual packets. So, $\pi^g$ transmits the packets with minimum speed $3W/D$ (which is also small if $D$ is large). 
Hence, the energy consumed by $\pi^g$ decreases as $D$ increases (other parameters being fixed). Figure \ref{fig:E_vs_D_multiFn} verifies this argument as 
it shows the plot of energy consumed as a function of peak AoI ($D$) when initial AoI $\Delta(0)=0$, size of packets is $W=1$ Mbit, time horizon $T=100$ msec, and inter-generation time of packets is uniformly distributed in interval $[0,3]$ msec.  
Note that as $D\to T$, energy consumption of $\pi^g$ converges to $0$ because fewer packets (transmitted at slower speed) are sufficient to satisfy the peak AoI constraint \eqref{eq:constraint}. 

\begin{figure} 
	\begin{subfigure} {0.49\textwidth}
	\begin{center}
		\includegraphics[width=0.9\linewidth]{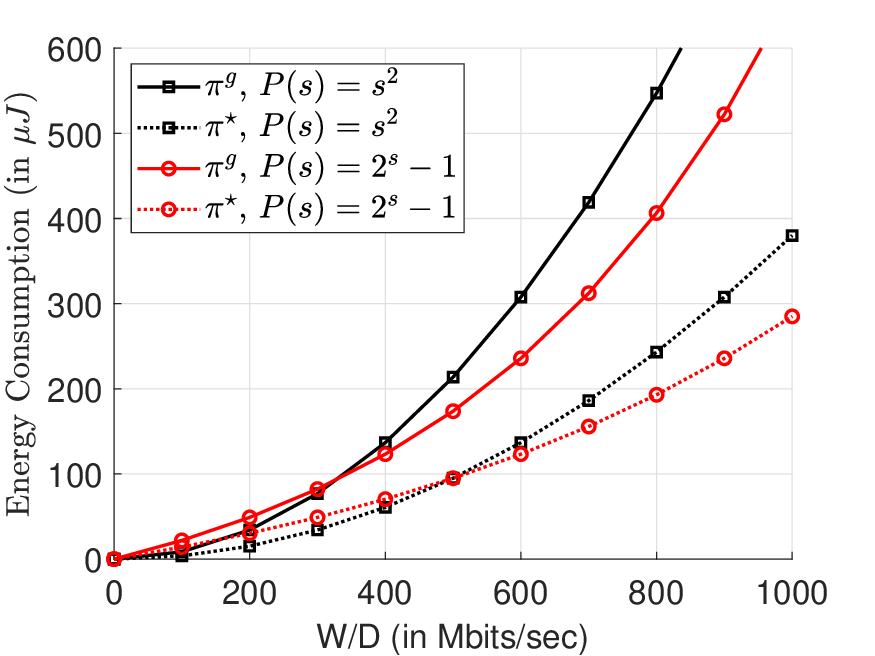} 
	\caption{\small{Energy consumed by $\pi^g$, and an optimal offline policy $\pi^\star$.}}
	\label{fig:E-W-multiFn}
	\end{center}
\end{subfigure}
\hspace{2ex}
\begin{subfigure}{0.49\textwidth}
	\begin{center} 
		\includegraphics[width=0.9\linewidth]{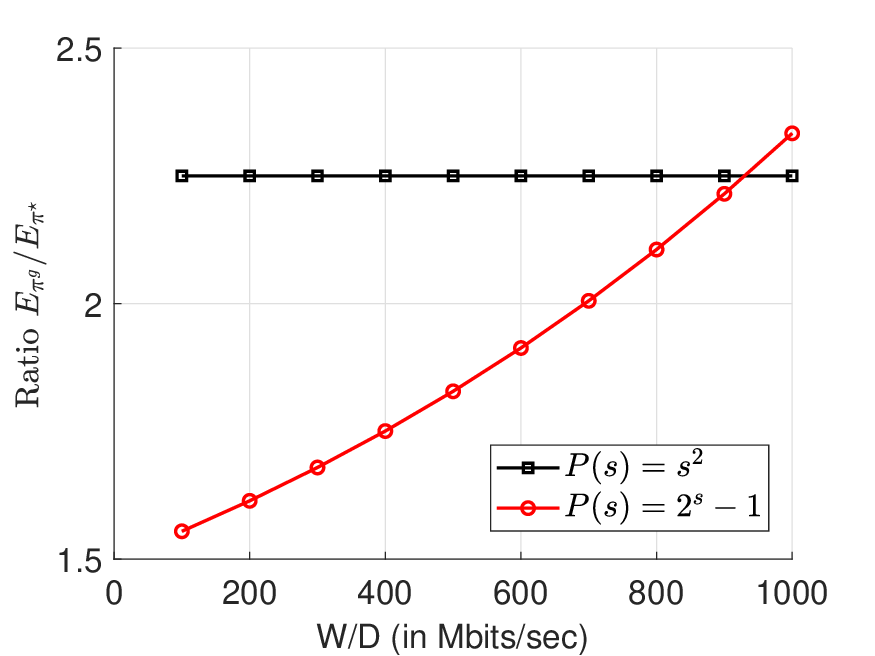} 
	\caption{\small{The ratio of the energy consumed by $\pi^g$, and an optimal offline policy $\pi^\star$.}}
	\label{fig:Ratio}
	\end{center}
\end{subfigure}
\caption{\small{Effect of the ratio $W/D$ on the performance of $\pi^g$.}}
\label{fig:vsW/D}
\end{figure}

Further, from Theorem \ref{thm:CR-bounds} (and Remark \ref{remark:cr-poly-exp}), we know that for exponential power functions $P(s)=2^s-1$, the competitive ratio of $\pi^g$ increases exponentially with increase in the ratio $W/D$. Also, for polynomial power function $P(s)=s^2$, the competitive ratio of $\pi^g$ is independent of the ratio $W/D$. To visualize this numerically, we consider a setting where the initial AoI $\Delta(0)=0$, peak AoI $D=5$ msec, time horizon $T=100$ msec, and the sequence of packet generation times is $\sigma=\{kD/3|\forall k\in \{1,2,\cdots\}\}\cup\{kD/2|\forall k\in\{1,2,\cdots\}\}$. In this setting, we simulate the proposed greedy policy $\pi^g$, and an optimal offline $\pi^\star$ (where $\pi^\star$ is found analytically, and its energy consumption matches the universal lower bound ($ULB$; Theorem \ref{thm:universal-lower-bound})), for different values of $W$ (i.e. $W/D$). 
Figure \ref{fig:E-W-multiFn} shows the plot of energy consumption of $\pi^g$ and $\pi^star$ for different values of $W/D$, for the power functions $P(s)=s^2$, and $P(s)=2^s-1$. Figure \ref{fig:Ratio} shows that the ratio of the energy consumed by $\pi^g$ and $\pi^\star$ increases exponentially with increase in $W/D$ when power function is $P(s)=2^s-1$. Also, when  the power function is $P(s)=s^2$, the ratio of the energy consumed by $\pi^g$ and $\pi^\star$ is independent of $W/D$.

Finally, we conclude this section by analyzing the effect of peak power/speed constraint \eqref{eq:p-speed-constraint} on the energy consumption. We consider a system with initial AoI $\Delta(0)=0$, peak AoI $D=3$ msec, and time horizon $T=100$ msec. We consider a set of 56 packet generation sequences $\sigma$, with inter-generation times that are either constant, or random (uniform/exponential/Rayleigh distribution), restricted in the interval $(0,2.75]$. For each packet generation sequence, we simulate the following four policies.
\paragraph{$\pi^g$ $(3W/D)$} 
Algorithm \ref{algo:greedy} with speed function \eqref{eq:speed} (i.e. $\max\{W/(d(t)-t),3W/D\}$).

\paragraph{$\pi^g$ $(4W/D)$} 
Algorithm \ref{algo:greedy} with modified speed function $\max\{W/(d(t)-t),4W/D\}$.

\paragraph{$\pi^g$ $(2W/D)$} 
Algorithm \ref{algo:greedy} with modified speed function $\max\{W/(d(t)-t),2W/D\}$.

\paragraph{$\pi^p$}
an interruptive policy, that at any time $t$, transmits the latest generated packet with remaining size (number of bits) at most $s^{\max}(d(t)-t)$, at maximum speed $s^{\max}$. Note that $\pi^p$ interrupts an ongoing transmission at time $t$, only if a newly generated packet can be completely transmitted until time $d(t)$ at maximum speed $s^{\max}$.

\begin{remark}
	For policies $(a)$, $(b)$ and $(c)$, whenever speed $s(t)$ exceeds $s^{\max}$, we limit it to $s^{\max}$. Hence, the speed constraint \eqref{eq:p-speed-constraint} is always satisfied, and we only need to consider the peak AoI constraint \eqref{eq:constraint} (equivalently, the deadline constraint \eqref{eq:p-deadline-constraint}).
\end{remark}

\begin{figure} 
	\begin{subfigure} {0.49\textwidth}
		\begin{center}
			\includegraphics[width=0.9\linewidth]{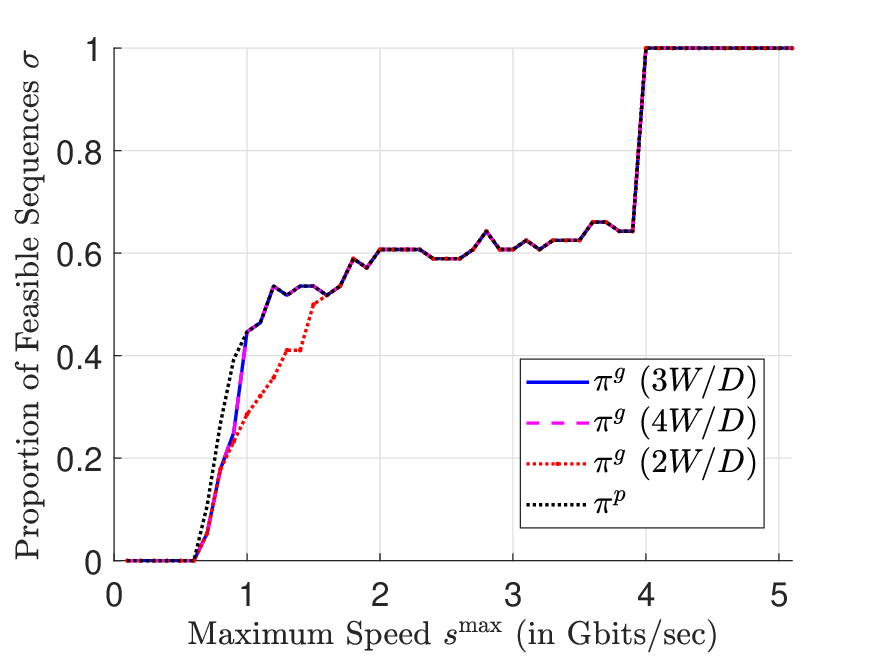} 
			\caption{\small{Proportion of the number of packet generation sequences for which the deadline constraint \eqref{eq:p-deadline-constraint} is satisfied, as a function of the maximum transmission speed $s^{\max}$.}}
			\label{fig:feasible_sequences}
		\end{center}
	\end{subfigure}
\hspace{2ex}
	\begin{subfigure}{0.49\textwidth}
		\begin{center} 
			\includegraphics[width=0.9\linewidth]{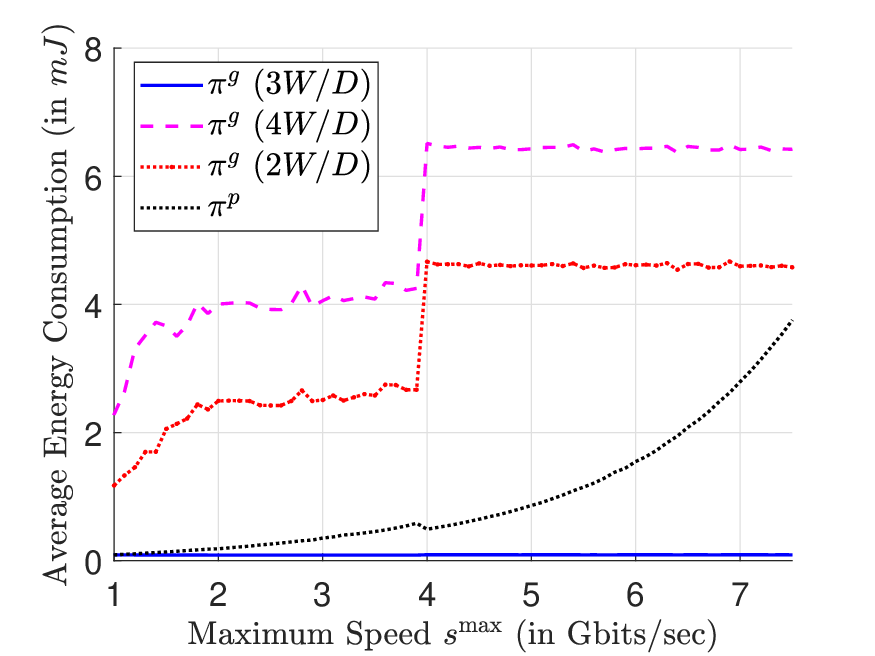} 
			\caption{\small{Energy consumption as a function of the maximum transmission speed $s^{\max}$}.}
			\label{fig:energy_vs_smax}
		\end{center}
	\end{subfigure}
	\caption{\small{Effect of the maximum speed constraint \eqref{eq:p-speed-constraint}.}}
	\label{fig:smax}
\end{figure}

For each of the four policies,  
Figure \ref{fig:feasible_sequences} plots the ratio of the number of packet generation sequences for which the deadline constraint \eqref{eq:p-deadline-constraint} is satisfied, and the total number of packet generation sequences (equal to $56$). Also, for power function $P(s)=2^s-1$, Figure \ref{fig:energy_vs_smax} plots the energy consumed by the policies for different values of $s^{\max}$, averaged across the packet generation sequences for which the deadline constraint \eqref{eq:p-deadline-constraint} is satisfied.

From Figure \ref{fig:feasible_sequences}, note that when $s^{\max}<2W/D\approx 0.67$ Gbits/sec, none of the policies could satisfy the deadline constraint \eqref{eq:p-deadline-constraint}, for any of the packet generation sequences. This is because $T>3D/2$, and as shown in Corollary \ref{cor:T<3D/2}, when $s^{\max}<2W/D$, and $T>3D/2$, the deadline constraint \eqref{eq:p-deadline-constraint} can never be satisfied. Further, when $s^{\max}\ge 3W/D=1$ Gbits/sec, the greedy policy $\pi^g$ ($3W/D$) satisfies the deadline constraint \eqref{eq:p-deadline-constraint} for maximum number of packet generation sequence. This is in accordance with Lemma \ref{lemma:smax>3W/D}, which shows that when $s^{\max}\ge 3W/D$, $\pi^g$ ($3W/D$) satisfies the deadline constraint \eqref{eq:p-deadline-constraint}for all packet generation sequences, for which the constraint \eqref{eq:p-deadline-constraint} can be satisfied by any causal policy. Moreover, in Figure \ref{fig:energy_vs_smax}, note that the energy consumed by $\pi^g$ $(3W/D)$ is minimum among all four policies. This justifies the choice of $3W/D$ in the speed function \eqref{eq:speed} (i.e. $\max\{W/(d(t)-t),3W/D\}$) of the greedy policy (Algorithm \ref{algo:greedy}).

	In Figures \ref{fig:feasible_sequences}, 
	note that at $s^{\max}=4$ Gbits/sec, there is a sudden increase in the number of sequences for which the deadline constraint \eqref{eq:p-deadline-constraint} is satisfied by the policies. To understand the reason, note that peak AoI $D=3$ msec, and 
	several packet generation sequences have inter-generation time equal to $2.75$ msec (maximum inter-generation time for the considered sequences). Whenever inter-generation time is $2.75$ msec, to satisfy the deadline constraint \eqref{eq:p-deadline-constraint}, policies need to transmit a complete packet of size $1$ Mbits in $0.25$ msec, which is possible only when $s^{\max}\ge 4$ Gbits/sec. Therefore, when $s^{\max}=4$ Gbits/sec, the deadline constraint \eqref{eq:p-deadline-constraint} becomes feasible for all the remaining packet generation sequences. 

\section{Conclusion} \label{sec:conclusion} 
In this paper, we considered the problem of minimizing energy consumption at a node, under a peak age of information (AoI) constraint. For this, we considered a speed scaling model with arbitrary packet arrival, where at each time instant, the node needs to decide which of the available packets to transmit (the node can discard a partially transmitted packet, and begin transmitting a newly arrived packet), and at what speed. We assumed that the power consumption is an increasing and convex function of transmission speed, and derived a lower bound on the competitive ratio (CR) of all causal policies. 
We showed that the CR of any causal policy depends on the power function, size of the packets, and the maximum allowed peak AoI. 
Then, we proposed a simple non-interruptive greedy policy, and bounded its CR by deriving some structural results for an optimal offline policy. Importantly, 
we showed that the CR of the proposed greedy policy has similar order of dependence on the system parameters (power function, size of the packets, and the maximum allowed AoI) as the derived lower bound on the CR of all causal policies.


\bibliographystyle{IEEEtran}
\bibliography{reflist}

\appendices

\newpage

%
	
{\LARGE \centerline{Online Energy Minimization Under a Peak Age of}} \vspace{-2ex}{\LARGE \centerline{Information Constraint}}
\vspace{-1ex}
{\large \centerline{Kumar Saurav and Rahul Vaze}}

{\Large \centerline{Supplementary Material}}
\vspace{-2ex}
\section{Proof of Lemma \ref{lemma:constant-speed-better}} \label{appendix:constant-speed-better}

Let a policy $\pi$ transmits $w$ bits in interval $[p,q)$ with constant speed $s_w(t)=w/(q-p)$. 
Also, consider a general policy $\pi'$ that transmits $w$ bits in interval $[p,q)$ with arbitrary speed $s(t)$ $\forall t\in [p,q)$.  
The energy consumed by $\pi$ is 
\begin{align}
	P\left(\frac{w}{q-p}\right)(q-p) &= P\left(\frac{\int_{t=p}^{q}s(t)dt}{q-p}\right)(q-p)\stackrel{(a)}{\le} \frac{\int_{t=p}^{q}P(s(t))dt}{q-p}(q-p)=\int_{t=p}^{q}P(s(t))dt,
\end{align}
where in $(a)$, we used Jensen's inequality ($P(\cdot)$ is a convex function). Therefore, $\pi$ consumes at most as much energy as any other policy $\pi'$ that transmits $w$ bits in interval $[p,q)$. 

Further, the minimum energy consumed in interval $[p,q)$ is $P(w/(q-p))(q-p)$ (equal to the energy consumed by $\pi$ in interval $[p,q)$).

\section{Proof of Corollary \ref{cor:P(w/y)y-decreasing-y}} \label{appendix:proof-cor:P(w/y)y-decreasing-y}
To prove Corollary \ref{cor:P(w/y)y-decreasing-y}, it is sufficient to show that for fixed $w$ and $y_1<y_2$, $P(w/y_1)y_1> P(w/y_2)y_2$. Without loss of generality, let a causal policy $\pi$ transmits $w$ bits in interval $[0,y_2)$. From Lemma \ref{lemma:constant-speed-better}, we know that minimum energy (equal to $P(w/y_2)y_2$) is consumed in interval $[0,y_2)$ if the $w$ bits are transmitted with constant speed (equal to $w/y_2)$) over the entire interval $[0,y_2)$. This implies that if another policy $\pi'$ transmits these $w$ bits over the interval $[0,y_1)$ (where $y_1<y_2$) with constant speed $w/y_1$ (and with zero speed over the interval $[y_1,y_2)$), it consumes energy $P(w/y_1)y_1> P(w/y_2)y_2$, because for $\pi'$, the speed is not constant in interval $[0,y_2)$. So, when $y_1<y_2$, $P(w/y_1)y_1> P(w/y_2)y_2$.


\section{Proof of Theorem \ref{thm:lb-CR}}
\label{appendix:proof-thm-lb-CR}
Consider the scenario, where the AoI at the monitor at time $t=0$ is $\Delta(0)=D/2$ (thus, $d(0)=D-\Delta(0)=D/2$), the time horizon $T=3D/2-\delta$ (for $\delta\to 0^+$), and the packets are being generated according to one of the following two instances of packet generation times $\sigma$: $(i)$ $\sigma_1=\{0,D/4,D/2\}$, and $(ii)$ $\sigma_2=\{0,D/4,5D/6\}$. An optimal offline policy $\pi^\star$ knows the actual instance $\sigma$ according to which the packets are being generated. Therefore, if the actual instance is $\sigma=\sigma_1$, then $\pi^\star$ would consume at most $P(2W/D)D$ units of energy. This is because $P(2W/D)D$ units of energy is sufficient to satisfy the deadline constraint \eqref{eq:deadline-constraint} by transmitting the packets generated at time $t=0$ and $t=D/2$ with constant speed $2W/D$ over the intervals $[0,D/2)$, and $[D/2,D)$ respectively. 
Similarly, if the actual instance $\sigma=\sigma_2$, then $\pi^\star$ consumes at most $P(12W/5D)5D/4$ units of energy, required to transmit the three packets generated at time $t=0$, $t=D/4$ and $t=5D/6$ with constant speed $12W/5D$ over the intervals $[0,5D/12)$, $[5D/12,5D/6)$ and $[5D/6,5D/4)$ respectively (because it satisfies the deadline constraint \eqref{eq:deadline-constraint}). 

However, a causal policy $\pi$ does not know the actual instance of packet generation times in advance. 
Therefore, until time $t=D/2$, $\pi$ cannot distinguish if the packets are being generated according to $\sigma_1$ or $\sigma_2$. Also, $\pi$ needs to deliver at least one of the packets generated in interval $[0,D/2)$ before time $t=d(0)=D/2$ (to satisfy the deadline constraint \eqref{eq:deadline-constraint}). Hence, consider the two possible cases:
    \paragraph{$\pi$ delivers packet $2$ (generated at time $t=D/4$) before $d(0)=D/2$} then, 
    $\pi$ consumes at least $P(4W/D)D/4$ units of energy in interval $[0,D/2)$, 
	required to transmit packet $2$ with constant speed $4W/D$, over the interval $[D/4,D/2)$. Also, the deadline at time $t=D/2$ will be reset to $5D/4$. 
	If the packets are generated according to $\sigma=\sigma_1$, $\pi$ would need to transmit packet 3 (generated at time $D/2$) before the new deadline, i.e., $5D/4$. Thus, $\pi$ would consume at least $P(4W/3D)3D/4$ units of energy in interval $[D/2,5D/4)$, required to transmit packet 3 
	with constant speed $4W/3D$. Since $\pi^\star$ consumes at most $P(2W/D)D$ units of energy (when $\sigma=\sigma_1$), the ratio of energy consumed by $\pi$ and $\pi^\star$ in interval $[0,T]$ is  
	\begin{align} \label{eq:ratio1}
		\frac{E_{\pi}}{E_{\pi^\star}}&\ge \frac{P(4W/D)D/4+P(4W/3D)3D/4}{P(2W/D)D}=\frac{P(4W/D)}{4P(2W/D)}+\frac{1}{2}.
	\end{align}
   \paragraph{$\pi$ delivers packet $1$, and transmits $\gamma W$ bits of packet $2$ (where $\gamma\in[0,1)$) until the initial deadline $d(0)=D/2$} then, energy consumed by $\pi$ in interval $[0,D/2)$ is at least $P(2(1+\gamma)W/D)D/2$, and there are following two possible sub-cases. 
   	
   	$(1)$ $\gamma\ge 0.07$: let $\sigma=\sigma_1$ (i.e., packet $3$ is generated at time $t=D/2$). 
   		
   		$(i)$ If $\pi$ interrupts the transmission of packet $2$, and begins to transmit packet $3$, still it ($\pi$) would consume at least $P(2W/D)D/2$ units of energy in interval $[D/2,T]$ (required to transmit packet $3$ over the interval $[D/2,D)$ with constant speed $2W/D$). So, the total energy consumed by $\pi$ in interval $[0,T]$ is $P(2(1+\gamma)W/D)D/2+P(2W/D)D/2\ge P(2.14W/D)D/2+P(2W/D)D/2$ (for $\gamma=0.07$).
   		
   		$(ii)$ If $\pi$ delivers packet $2$ at any time $t=\tau_2<D$, still it would need to transmit packet $3$ (because at $\tau_2$, the deadline would be $d_2=t_2+D=D/4+D<T$). Therefore, in this case, if $\pi$ transmits packet $3$ over the interval $[\tau_2,d_2)$, then the total energy that $\pi$ would consume in interval $[0,T]$ would be 
   		\begin{align} \label{eq:min-energy-lbCP-case2-2}
   			E_{\pi}\ge P&\left(\frac{(1+\gamma)W}{D/2}\right)\frac{D}{2}+P\left(\frac{(1-\gamma)W}{\tau_2-D/2}\right)\left(\tau_2-\frac{D}{2}\right) +P\left(\frac{W}{5D/4-\tau_2}\right)\left(\frac{5D}{4}-\tau_2\right).
   		\end{align} 
   	    Minimizing \eqref{eq:min-energy-lbCP-case2-2} with respect to $\gamma\ge 0.07$ and $\tau_2\in[D/2,d_1)$ (where $d_1=0+D$), we get
   	    $E_{\pi}\ge P(12W/5D)(5D/4)$ (with equality if $\gamma=0.2$ and $\tau_2=5D/6$).
    
    Therefore, the ratio of energy consumed by $\pi$ and $\pi^\star$ in interval $[0,T]$ is
    \begin{align} \label{eq:ratio2}
    	\frac{E_{\pi}}{E_{\pi^\star}}&\ge \min\bigg\{\frac{P(2.14W/D)D/2+P(2W/D)D/2}{P(2W/D)(D/2)}, \frac{P(2.4W/D)(5D/4)\}}{P(2W/D)(D/2)}\bigg\}, \nonumber \\
    	&=\min\bigg\{\frac{P(2.14W/D)}{P(2W/D)}+1, \ \ \frac{2.5P(2.4W/D)}{P(2W/D)}\bigg\}.
    \end{align}

   	$(2)$ $\gamma<0.07$: let $\sigma=\sigma_2$ (i.e., packet $3$ is generated at time $t=5D/6$). 
   	
   	$(i)$ If $\pi$ interrupts the transmission of packet $2$ (does not deliver packet $2$), then it must deliver packet $3$ before $d_1=0+D$. Therefore, $\pi$ consumes at least $P(W/(D-5D/6))(D-5D/6)$ units of energy in interval $[D/2,T]$ (required to transmit packet $3$ over the interval $[5D/6,D)$, with constant speed $6W/D$). Thus, total energy consumed in interval $[0,T]$ is 
   	$E_{\pi}\ge P(2(1+\gamma)W/D)(D/2)+P(6W/D)(D/6)\ge P(2W/D)(D/2)+P(6W/D)(D/6)$ (for $\gamma=0$).
   	
   	$(ii)$ If $\pi$ delivers packet $2$ at any time $t=\tau_2$ (since only $\gamma W <0.07W$ bits of packet 2 are transmitted until time $D/2$, $\tau_2>D/2$), followed by packet $3$ (deadline constraint \eqref{eq:deadline-constraint} cannot be satisfied in interval $[D,T]$ without transmitting packet $3$ (generated at time $t=5D/6$). Therefore, the energy consumed by $\pi$ in interval $[D/2,T)$ is at least $P((1+0.93)W/(5D/4-D/2))$, required to transmit the remaining bits of packet $2$ (which is at least $0.93W$), and $W$ bits of packet $3$ over the interval $[D/2,5D/4)$ (note that $d_2=t_2+D=D/4+D=5D/4$, and hence, to satisfy the deadline constraint \eqref{eq:deadline-constraint}, $\pi$ must transmit the $W$ bits of packet $3$ before $5D/4$). 
   So, total energy consumed by $\pi$ in interval $[0,T]$ is $E_{\pi}\ge P(2(1+\gamma)W/D)(D/2)+P((1+0.93)W/(5D/4-D/2))(5D/4-D/2)\ge P(2W/D)(D/2)+P(7.72W/3D)(3D/4)$ (for $\gamma=0$).
   
   Therefore, the ratio of energy consumed by $\pi$ and $\pi^\star$ in interval $[0,T]$ is
   \begin{align} \label{eq:ratio3}
   	\frac{E_{\pi}}{E_{\pi^\star}}&\ge \min\bigg\{\frac{P(2W/D)(D/2)+P(6W/D)(D/6)}{P(12W/5D)(5D/4)}, \frac{P(2W/D)(D/2)+P(7.72W/3D)(3D/4)}{P(12W/5D)(5D/4)}\bigg\}, \nonumber \\
   	&\ge \min\bigg\{\frac{2P(6W/D)}{15P(2.4W/D)}, \ \ \frac{3P(2.57W/D)}{5P(2.4W/D)}\bigg\}.
   \end{align}

From \eqref{eq:ratio1}, \eqref{eq:ratio2} and \eqref{eq:ratio3}, we conclude that 
\begin{align}
	\textsc{cr}_{\pi}\ge \frac{E_{\pi}}{E_{\pi^\star}} \ge \frac{c_1P(c_2W/D)}{P(c_3W/D)},
\end{align}
where $c_1$, $c_2$ and $c_3$ are finite positive constants, $c_1\ge (2/15)>0.13$, $c_2-c_3\ge 0.14$, and $c_2/c_3\ge 1.07$. 

\section{} \label{app:AAoI}
The following example shows that in the optimization problem \eqref{eq:objective-init}--\eqref{eq:constraint}, instead of peak AoI constraint, if we consider a constraint on the maximum average AoI, then the competitive ratio of all causal policies will be infinite.
\begin{example} \label{ex:aaoi-cr-infty}
	Let AoI at time $t=0$ be $\Delta(0)=0$, and the time horizon $T=4$ time units. Consider average AoI constraint that requires that the average AoI in interval $[0,T]$ should be less than $D=1.25+\epsilon$, where $\epsilon\to 0^+$. Packets are generated according to one of the two sequences of packet generation times $\sigma$: $(i)$ $\sigma_1=\{1\}$, and $(ii)$ $\sigma_2=\{1,2\}$. 
	
	Note that in case of $\sigma_1$, the average AoI constraint can be satisfied only if packet $1$ (the packet generated at time $t=1$) is completely transmitted until time $t=1+\epsilon$, which would consume at least $P(W/\epsilon)\epsilon\to\infty$ units of energy. On the other hand, in case of $\sigma_2$, the average AoI constraint can be satisfied by transmitting packet $1$ (generated at time $t=1$) at a constant speed of $W/1$ over the interval $[1,2)$, followed by packet $2$ (generated at time $t=2$) at a constant speed of $W/1$ over the interval $[2,3)$, consuming a total of $2P(W/1)<\infty$ units of energy. 
	Until time $t=1+\epsilon$, since a causal policy does not know the actual sequence of packet generation times $\sigma$, we may have two types of causal policies: 
	
	$(i)$ Causal policies $\pi$ that completely transmit packet $1$ until time $t=1+\epsilon$, (consuming at least $P(W/\epsilon)\epsilon\to\infty$ units of energy). If $\sigma=\sigma_2$, then an optimal offline policy $\pi^\star$ would consume at most $2P(W/1)<\infty$ units of energy, which implies that the competitive ratio of $\pi$ is infinite.
	$(ii)$ Causal policies $\pi'$ that do not completely transmit packet $1$ until time $t=1+\epsilon$. If $\sigma=\sigma_1$, then $\pi'$ will not satisfy the average AoI constraint, and hence, will be infeasible.
	
	Therefore, if there is an average AoI constraint (instead of peak AoI constraint), then the competitive ratio is unbounded for all causal policies. 
\end{example}

\section{Proof of Proposition \ref{prop:large-X}} \label{appendix:prop-large-X}
Consider time $t$, when $\pi^g$ begins to transmit a packet $j$ with speed $s^g(t)>3W/D$. To prove Proposition \ref{prop:large-X}, we need to show that no packet must have been generated in interval $[t-2D/3,t)$. We show this following the method of contradiction. 
Let at least one packet be generated in the interval  $[t-2D/3,t)$. Since $\pi^g$ never idles when there is a fresh packet to transmit, we must have one of the following two cases:
\paragraph{$\pi^g$ begins to transmit a packet $i$ generated in interval  $[t-2D/3,t)$, at some time $r_i<t$}  
By hypothesis, $\pi^g$ begins to transmit packet $j$ at time $t$.   
Since $\pi^g$ is a non-interruptive policy (Remark \ref{remark:non-preemptive}), this is possible only if until time $t$, $\pi^g$ completely transmits packet $i$ that it began to transmit at time $r_i<t$. Therefore, let $\pi^g$ completely transmits packet $i$ until time $t$. Since generation time of packet $i$ is $g_i\ge t-2D/3$, this implies that the deadline at time $t$ is $d(t)\ge g_i+D\ge t+D/3$, which further implies that $d(t)-t\ge 3W/D$. But, by definition of $\pi^g$, at any time $t$, when $\pi^g$ begins to transmit a packet, if $d(t)-t\ge D/3$, $\pi^g$ begins to transmit the packet at speed $3W/D$. This contradicts the hypothesis that at time $t$, $\pi^g$ begins to transmit packet $j$ at speed greater than $3W/D$. 
Hence, this case is not possible. 

\paragraph{In interval $[t-2D/3,t)$, $\pi^g$ remains busy transmitting packets that were generated before time $t-2D/3$} 
Such an event cannot happen. 
To show this, let packet $i$ be the latest packet generated before time $t-2D/3$ (say, at time $t_i=t-2D/3-\delta$, where $\delta>0$). Because $\pi^g$ only transmits a fresh packet, in interval $[t-2D/3,t)$, $\pi^g$ may transmit at most two packets generated before $t-2D/3$\ \ : $(i)$ a packet $\hat{i}$ that was being transmitted by $\pi^g$ when packet $i$ was generated at time $t_i$, and $(ii)$ packet $i$ itself. Since $\pi^g$ transmits a packet with speed at least $3W/D$ (Eq.\eqref{eq:speed}), it takes at most $D/3$ time units to completely transmit (deliver) a packet. Therefore, $\pi^g$ would finish transmitting both packet $\hat{i}$ and packet $i$ before $t_i+2D/3=t-\delta$, and hence, must begin to transmit a packet generated in interval $[t-2D/3,t)$ in sub-interval $[t-\delta,t)$, thus proving that this case is not possible. 

Thus, we conclude that none of the above two cases are possible. This contradicts the assumption that a packet was generated in interval $[t-2D/3,t)$.

\section{Proof of Corollary \ref{cor:large-speed-immediate-transmission}} \label{appendix:proof-cor-large-speed-immediate-transmission}
Let transmission of packet $j$ begins at time $t$. So, $t_j\le t$, where $t_j$ is the generation time of packet $j$. From Proposition \ref{prop:large-X}, we know that no packet is generated in interval $[t-2D/3,t)$. Also, it has been shown in proof of Proposition \ref{prop:large-X} that the transmission of packets generated before $t-2D/3$ (that are transmitted by $\pi^g$) gets completed before time $t$, and hence, $t_j$ cannot be less than $t-2D/3$ (because transmission of packet $j$ begins at time $t$). Hence, we must have $t_j=t$. 

Further, $\pi^g$ transmits packet $j$ with constant speed $W/(d(t)-t)=W/(d(t_j)-t_j)$ starting time $t=t_j$. So, transmission of packet $j$ completes at time $d(t_j)$.

\section{Proof of Proposition \ref{prop:facts-min-energy-opt-frame}} \label{appendix:proof-prop-facts-min-energy-opt-frame}
\paragraph{Proof of Property 1}
Let $\pi^\star$ transmits the $W$ bits of a packet $i$ with speed that varies with time. Now, consider another policy $\pi'$, identical to $\pi^\star$, except that it ($\pi'$) transmits the $W$ bits of packet $i$ with constant speed over the same time-interval where $\pi^\star$ transmits packet $i$. Due to convexity of power function $P(\cdot)$, we know that over any given interval of time, transmitting a packet with constant speed consumes minimum energy. 
Therefore, $\pi'$ consumes less energy than $\pi^\star$. But this cannot be true, because $\pi^\star$ is an optimal offline policy. Hence, $\pi^\star$ must transmit the $W$ bits of each packet $i$ with constant speed. 

\paragraph{Proof of Property 2}
Since $\pi^\star$ knows the generation time of all the packets in advance, it never transmits any packet partially (because transmitting a packet partially consumes energy, without meeting the deadline constraint \eqref{eq:deadline-constraint}). Also, $\pi^\star$ only transmits fresh packets (Remark \ref{remark:only-fresh-packet}). Therefore, it never interrupts transmission of any packet to transmit it later. 
Hence, $\pi^\star$ never interrupts any ongoing packet transmission.

\paragraph{Proof of Property 3}

Note that when $\pi^\star$ begins to transmit packet $i+1$ at time $r_{i+1}$, the deadline is $d_i$ (the latest delivered packet at time $r_{i+1}$ is packet $i$). Therefore, $\tau_{i+1}<d_i$, where $\tau_{i+1}$ is the time when packet $i+1$ is delivered. 
Therefore, packet $i+1$ is delivered in one of the frames $0$ to $i$. Hence, for $i=0$, the only possibility is that packet 1 is delivered in frame 0. Next, using induction, we show that for all $i$, $\pi^g$ delivers packet $i+1$ in frame $i$. Let packet $i$ is delivered in frame $i-1$. 
Then, there are two possible cases: $(i)$ packet $i+1$ is delivered in frame $i$, and $(ii)$ packet $i+1$ is delivered in frame $i-1$ (since packet $i$ is delivered in frame $i-1$, packet $i+1$ cannot have been delivered in a frame previous to frame $i-1$). 
Note that if packet $i+1$ is delivered in frame $i-1$, then there would be two packets (packet $i$ and packet $i+1$) that are delivered in frame $i-1$, at time $\tau_{i}$ and $\tau_{i+1}$ respectively. However, in this case, transmission of packet $i$ will be redundant because at the start of frame $i-1$ (i.e., time $d_{i-2}$), the deadline is $d_{i-1}$, and due to Proposition \ref{prop:feasible}, we know that delivery of a single packet is sufficient in interval $[d_{i-2},d_{i-1})$. 
Hence, $\pi^\star$ being an optimal offline policy, will not waste energy delivering both packets $i$ and $i+1$ in frame $i-1$. Thus, we conclude that for all $i$, $\pi^\star$ delivers packet $i+1$ in frame $i$. 

\paragraph{Proof of Property 4}
From Property 3, we know that exactly one packet (packet $i+1$) is delivered in frame $i$. Also, packet $i+1$ is transmitted with constant speed (Property 1). Therefore, transmission speed may decrease only after packet $i+1$ is delivered. However, $\pi^\star$ being an optimal offline policy, instead of 
delivering packet $i+1$ before the deadline $d_i$ (end of frame $i$), and decreasing the speed, it ($\pi^\star$) would transmit packet $i+1$ itself at lesser speed, over a larger interval. 
This follows due to convexity of power function $P(\cdot)$. 

\section{Proof of Theorem \ref{thm:universal-lower-bound}} \label{appendix:universal-lower-bound}
In each period $i$, an optimal offline policy $\pi^\star$ transmits at least $2W$ bits (Lemma \ref{lemma:2W-per-D}). Since length of each period is $D$, minimum energy is consumed in a period if $\pi^\star$ transmits the $2W$ bits with constant speed $2W/D$ throughout the period. Also, periods are overlapping. Therefore, if $t_1$ is the generation time of first packet that is transmitted by $\pi^\star$ in interval $[0,T]$, and $\tau_{\ell}$ is the time at which the last packet transmitted by $\pi^\star$ before $T$ is delivered to the monitor, then minimum energy is consumed by $\pi^\star$ in interval $[0,T]$ if $\pi^\star$ transmits at a constant speed of $2W/D$ in the interval $[t_1,\tau_{\ell})$. 
	However, note that the initial deadline $d(0)=D-\Delta(0)\le D$. Therefore, packet 1 must be delivered by time $D$. Also, the generation time of last packet must be greater than $T-D$ (so that $d_{\ell}>T$).
	Therefore, minimum energy is consumed while satisfying the deadline constraint \eqref{eq:deadline-constraint} if $t_1=D/2$, and $\tau_\ell=T-D/2$. 
	Moreover, the minimum energy consumed is $\max\{0,P(2W/D)(T-D)\}$, 
	which is a lower bound on the energy consumed by any feasible policy $\pi$ in interval $[0,T]$, for any instance of packet generation times.
	
\section{} 
\label{appendix:lemma-partitionAB}
In this section, let the packets be indexed in the order they are generated, irrespective of whether they are transmitted by $\pi^\star$ or not (because here we may need to consider packets that are transmitted by $\pi^g$ but not by $\pi^\star$). 
Also, in this section, let the deadline $d(t)$ at any time $t$ be defined according to $\pi^g$. \footnote{At any time $t$, the deadline $d(t)$ depends on the packets transmitted until time $t$. Therefore, $d(t)$ depends on the transmission policy.} 

Let $\pi^g$ transmits a packet $j$ with speed greater than $3W/D$. Then, from Corollary \ref{cor:large-speed-immediate-transmission}, we know that the transmission of packet $j$ must have begun at time $t_j$ (where $t_j$ is the generation time of packet $j$), and got completed at time $d(t_j)$. 
Therefore, in interval $[t_j,d(t_j))$, $\pi^g$ transmits packet $j$ with constant speed $W/(d(t_j)-t_j)$. Lemma \ref{lemma:partitionAB} shows that the optimal offline policy $\pi^\star$ also transmits an entire packet in interval $[t_j,d(t_j))$.  
\begin{lemma} \label{lemma:partitionAB}
	For each packet $j$ ($W$ bits) transmitted by $\pi^g$ in interval $[t_j,d(t_j))$ with constant speed $s^g(t)=W/(d(t_j)-t_j)>3W/D$, there exists a packet $\hat{j}$ such that $\pi^\star$ transmits the $W$ bits of packet $\hat{j}$ in interval $[t_j,d(t_j))$, with constant speed at least equal to $W/(d(t_j)-t_j)$.
\end{lemma}
\begin{IEEEproof}
	Let $\pi^g$ transmits a packet $j$ with speed greater than $3W/D$. From Proposition \ref{prop:large-X}, it follows that no packet is generated in interval $[t_j-2D/3,t_j)$, where $t_j$ is the generation time of packet $j$. Without loss of generality, let packet $\ell$ be the latest packet that is generated before time $t_j-2D/3$. 
	Since no packet is generated in interval $(t_\ell,t_j)$, packet $\ell$ remains fresh until time $t_j$. Also, note that $\pi^g$ takes at most $D/3$ time units to deliver $W$ bits (because $s^g(t)\ge 3W/D$), and $t_j-t_\ell>2D/3$. Therefore, even if $\pi^g$ was transmitting a previous packet when packet $\ell$ was generated at time $t_\ell$, it ($\pi^g$) delivers packet $\ell$ before time $t_j$. Therefore, the deadline at time $t_j$ for $\pi^g$ is $d(t_j)=d_{\ell}$. 
	
	Since no packet is generated in interval $(t_\ell,t_j)$, the generation time of the latest packet delivered by $\pi^\star$ until time $t_j$ can at most be $t_\ell$. Hence, the deadline for $\pi^\star$ at time $t_j$ is at most equal to $d_\ell$. Therefore, $\pi^\star$ must deliver a packet 
	in interval 
	$[t_j,d_\ell)$ to be feasible (Proposition \ref{prop:feasible}). Since $d_\ell=d(t_j)$, and the only packet available at time $t_j$ is packet $j$, 
	$\pi^\star$ must either transmit the $W$ bits of packet $j$ in interval $[t_j,d(t_j))$, or if some other packet $j'$ is generated in interval $(t_j,d(t_j))$, then transmit the $W$ bits of that packet in interval $[t_j,d(t_j))$. 
	Also, to minimize energy consumption, $\pi^\star$ must transmit the $W$ bits at constant speed (Proposition \ref{prop:facts-min-energy-opt-frame}, Property 1). 
	Therefore, we conclude that $\pi^\star$ transmits the $W$ bits of packet $\hat{j}$ (where $\hat{j}$ is either equal to $j$, or $j'$) 
	in interval $[t_j,d(t_j))$, with constant speed at least equal to $W/(d(t_j)-t_j)$.
\end{IEEEproof} 

\section{} \label{appendix:example_speed_constraint}
In the following example, we illustrate that 
under the speed constraint \eqref{eq:p-speed-constraint}, it may be possible to satisfy the deadline constraint \eqref{eq:p-deadline-constraint} for a sequence of packets with some constant inter-generation time $a$, but not for a sequence of packets with constant inter-generation time $a'<a$. 
This shows that the general intuition that smaller inter-generation times are better for AoI minimization is not correct in case of peak AoI constraint. 
\begin{example} \label{ex:small-not-better}
	Let $s^{\max}\in (2W/D,3W/D)$. Define $a=W/s^{\max}$ (minimum time required to completely transmit any packet), and $b=D-2a$. Since $s^{\max}\in (2W/D,3W/D)$, we have $a\in (D/3,D/2)$ and $b\in(0,a)$.
	Consider the scenario where the AoI at time $t=0$ is $\Delta(0)=a$ (i.e., the initial deadline $d(0)=D-\Delta(0)=a+b$), and the time horizon $T\to\infty$. A packet is generated at time $t=0$, and subsequently, packets are generated with constant inter-generation time $x$, i.e., the packet generation sequence is $\sigma=\{0,x,2x,\cdots\}$. 
	
	Note that when $x=a$, i.e., $\sigma=\{0,a,2a,\cdots\}$, a non-interruptive policy $\pi$ that at any time, transmits the latest generated packet at constant speed $s^{\max}$, satisfies the constraints \eqref{eq:p-deadline-constraint} and \eqref{eq:p-speed-constraint} (for any $i\in\bbN$, the packet generated at time $ia$ is completely transmitted until time $(i+1)a$). 
	In contrast, when $x=a-\epsilon$, where $\epsilon\in(0,a-b)$, any causal/offline policy that satisfies the speed constraint \eqref{eq:p-speed-constraint}, violates the deadline constraint \eqref{eq:p-deadline-constraint}, as discussed next.
	
	Consider the packet generation sequence $\sigma=\{0,x,2x,\cdots\}$ with fixed inter-generation time $x=a-\epsilon<a$, and any offline/causal policy $\pi'$ that satisfies the speed constraint \eqref{eq:p-speed-constraint} for input $\sigma$. Due to the speed constraint \eqref{eq:p-speed-constraint}, $\pi'$ takes at least $a$ time units to completely transmit a packet. Since $b<a$, in interval $[0,d(0))$ (i.e. $[0,a+b)$), only two packets are generated: at time $t=0$ and $t=a-\epsilon$ respectively. To satisfy the deadline constraint \eqref{eq:p-deadline-constraint}, at least one of these two packets must be completely transmitted by time $d(0)=a+b$. Since $\epsilon<a-b$, and the minimum time required to completely transmit a packet is $a$, until time $d(0)$, the only packet that can be completely transmitted by $\pi'$ is the packet generated at time $t=0$. Let the transmission of this packet (generated at time $t=0$) complete at time $\tau_0\ge a$. The deadline at time $t=\tau_0$ is $d(\tau_0)=0+D=2a+b$. In interval $[\tau_0,d(\tau_0))$, only the packet generated at time $a-\epsilon$ can be completely transmitted, by time $\tau_1\ge 2a$. The deadline at time $d(\tau_1)=3a+b-\epsilon$. Following the same argument, we get that for all $i\in\{1,2,\cdots,m\}$, where $m$ is the smallest integer greater than $b/\epsilon$, $\pi'$ finishes transmitting the $i^{th}$ packet (generated at time $i(a-\epsilon)$), at time $\tau_i\ge (i+1)a$, and the deadline at time $\tau_i$ is $d(\tau_i)=(i+2)a+b-i\epsilon$. Subsequently, to satisfy the deadline constraint \eqref{eq:p-deadline-constraint}, $\pi'$ must completely transmit the $m+1^{st}$ packet generated at time $(m+1)(a-\epsilon)$ in interval $[(m+1)a,(m+2)a+b-m\epsilon)$. However, the length of interval $[(m+1)a,(m+2)a+b-m\epsilon)$ is less than $a$ (because $m> b/\epsilon$), and hence, $\pi'$ cannot transmit any packet  completely in this interval. Therefore, $\pi'$ (i.e. any causal/offline policy that satisfies the speed constraint \eqref{eq:p-speed-constraint}) will violate the deadline constraint \eqref{eq:p-deadline-constraint} at time $d(\tau_m)=(m+2)a+b-m\epsilon$.
\end{example}

\section{Proof of Lemma \ref{lemma:no-feasible-policy}} \label{appendix:proof-lemma-no-feasible-policy}
To prove Lemma \ref{lemma:no-feasible-policy}, we consider the cases $s^{\max}=2W/D$ and $s^{\max}\in(2W/D,3W/D)$ separately. 

\subsection{$s^{\max}=2W/D$}
Let the initial deadline $d(0)=2D/3$, and the time horizon $T\to\infty$.
Also, let the packets be generated as per one of the following sequences of packet generation times $\sigma$: i) $\sigma_1=\{0,D/3\}\cup\{0+k(D/2)|k=1,2,\cdots,\infty\}$, and ii) $\sigma_2=\{0,D/3\}\cup\{(D/3)+k(D/2)|k=1,2,\cdots,\infty\}$. 

Note that for $\sigma_1$, any policy $\pi$ can satisfy both the constraints \eqref{eq:p-deadline-constraint} and \eqref{eq:p-speed-constraint} if and only if it completely transmits the packet generated at time $kD/2$, $\forall k=0,1,2,\cdots$, over the interval $[kD/2,(k+1)D/2)$ at maximum speed $s^{\max}=2W/D$. Similarly, for $\sigma_2$, the constraints \eqref{eq:p-deadline-constraint} and \eqref{eq:p-speed-constraint} are satisfied if (and only if) policy $\pi$ completely transmits the packets generated at time $D/3$ and $D/3+kD/2$, $\forall k=1,2,\cdots$, respectively over the intervals $[D/3,5D/6)$ and $[(D/3)+kD/2,(D/3)+(k+1)D/2)$, at maximum speed $s^{\max}=2W/D$. Hence, $\sigma_1,\sigma_2\in\Sigma(s^{\max})$.

Since a causal policy $\pi$ does not know the packet generation times in advance, until time $D/2$, it cannot differentiate between 
$\sigma_1$ or $\sigma_2$. Also, policy  $\pi$ cannot simultaneously transmit the packets generated at time $t=0$ and $t=D/3$ over the overlapping intervals $[0,D/2)$ and $[D/3,5D/6)$, respectively. Hence, if $\pi$ transmits the packet generated at time $t=0$ over the interval $[0,D/2)$, and $\sigma=\sigma_2$, then the deadline constraint \eqref{eq:p-deadline-constraint} will be violated. Similarly, if $\pi$ transmits the packet generated at time $t=D/3$ over the interval $[D/3,5D/6)$, and $\sigma=\sigma_1$, then also the the deadline constraint \eqref{eq:p-deadline-constraint} will be violated.


\subsection{$s^{\max}\in(2W/D,3W/D)$}
As in Example \ref{ex:small-not-better}, define $a=W/s^{\max}$ and $b=D-2a$ (where $D>0$). 
By definition, for any causal policy that satisfies the maximum speed constraint \eqref{eq:p-speed-constraint}, the minimum time required to completely transmit a packet is $a$ time units. Also, $D=2a+b$.
Note that for any $s^{\max}\in(2W/D,3W/D)$, we have $a\in(D/3,D/2)$ and $b\in(0,D/3]$, which implies $0<b<a$. 

Let $\epsilon>0$ be such that $\epsilon\le\min\{b/2,(a-b)/2\}$, and for some positive integer $m$, $b= (m+1)\epsilon$. 
Consider the scenario where the initial deadline $d(0)=a+\epsilon$, and the time horizon $T\to\infty$. The packets are generated as per one of the following two sequences of packet generation times $\sigma$: $(i)$ $\sigma_1=\{0,\epsilon\}\cup\{(a+b+\epsilon)+ja|j=0,1,2,\cdots,\infty\}$, and $\sigma_2=\{0,\epsilon\}\cup\{a+j(a-\epsilon)|j=0,1,2,\cdots,m\}\cup\{a+m(a-\epsilon)+ka|k=1,2,\cdots,\infty\}$.

For $\sigma=\sigma_1$, a policy $\pi(\sigma_1)$ can satisfy the constraints \eqref{eq:p-deadline-constraint} and \eqref{eq:p-speed-constraint} by transmitting i) the packet generated at time $t=\epsilon$, at constant speed $s^{\max}$, over the interval $[\epsilon,a+\epsilon)$, and ii) the packets generated at time $(a+b+\epsilon)+ja$, $\forall j=0,1,\cdots,\infty$, at constant speed $s^{\max}$, over the intervals $[(a+b+\epsilon)+ja,(a+b+\epsilon)+(j+1)a)$. Similarly, for $\sigma=\sigma_2$, a policy $\pi$ can satisfy the constraints \eqref{eq:p-deadline-constraint} and \eqref{eq:p-speed-constraint} by transmitting i) the packet generated at time $t=0$, at constant speed $s^{\max}$, over interval $[0,a)$, ii) the packets generated at time $t=a+j(a-\epsilon)$ ($\forall j=0,1,\cdots,\infty$), at constant speed $s^{\max}$, over the intervals $[(j+1)a, (j+2)a)$, and iii) the packets generated at time $t=a+m(a-\epsilon)+ka$  ($\forall k=1,\cdots,\infty$), at constant speed $s^{\max}$, over the intervals $[(m+1)a+ka,(m+2)a+ka)$. Hence, $\sigma_1,\sigma_2\in\Sigma(s^{\max})$.

Note that a causal policy $\pi$ does not know the packet generation times in advance, and hence, until time $t=a$, $\pi$ cannot distinguish whether the packets are being generated as per sequence $\sigma_1$ or $\sigma_2$. Also, in interval $[0,d(0))$ (where $d(0)=a+\epsilon$ is the initial deadline), $\pi$ must completely transmit either the packet generated at time $t=0$, or the packet generated at time $t=\epsilon$. Therefore, we have two cases. 
\paragraph{In interval $[0,a+\epsilon)$, $\pi$ completely transmit the packet generated at time $t=0$} Let $\pi$ finish transmitting the packet generated at time $t=0$, at time $t=\tau_0$. Since the time required to completely transmit a packet is $a$, we have $\tau_0\ge a$. The deadline at time $t=\tau_0$ is $\d(\tau_0)=D=2a+b$. In interval $[\tau_0,2a+b)$, only the packet generated at time $t=\epsilon$ can be completely transmitted. Let $\pi$ finish transmitting the packet generated at time $t=\epsilon$, at time $t=\tau_1\ge 2a$. Then, the deadline at time $t=\tau_1$ becomes $d(\tau_1)=2a+b+\epsilon$. To satisfy the deadline constraint \eqref{eq:p-deadline-constraint}, a new packet must be completely transmitted in interval $[\tau_1,d(\tau_1))$. However, since $\tau_1\ge 2a$, and $d(\tau_1)=2a+b+\epsilon$ (where $\epsilon<a-b$), the length of interval $[\tau_1,d(\tau_1))$ is less than $a$. Therefore, $\pi$ cannot transmit any packet completely in interval $[\tau_1,d(\tau_1))$. Hence, the deadline constraint will be violated at time $t=d(\tau_1)$.
 
\paragraph{In interval $[0,a+\epsilon)$, $\pi$ completely transmit the packet generated at time $t=\epsilon$} This is possible only if $\pi$ transmits the packet generated at time $t=\epsilon$ at constant speed $s^{\max}$, over the interval $[\epsilon,a+\epsilon)$. Hence, $\pi$ cannot begin to transmit the packet generated at time $t=a$, until time $t= a+\epsilon$. 
Thus, following the arguments in  Example \ref{ex:small-not-better}, we get that while transmitting the packets generated at times $\{a+j(a-\epsilon)|j=0,1,\cdots,m\}$, policy $\pi$ will violate the deadline constraint \eqref{eq:p-deadline-constraint} at time $t=(m+2)a$. 

Hence, for any causal policy $\pi$ that satisfies the speed constraint \eqref{eq:p-speed-constraint}, the deadline constraint \eqref{eq:p-deadline-constraint} is violated either for $\sigma=\sigma_1$, or $\sigma=\sigma_2$.

\section{Proof of Theorem \ref{thm:arbitrary-packet-size}} \label{appendix:proof-thm:arbitrary-packet-size}
    $(1)$ The first result follows directly from Theorem \ref{thm:CR-bounds}, assuming that the packets transmitted by the offline optimal policy $\pi^\star$ has minimum size $w$, while the packets transmitted by the greedy policy $\pi^g$ has maximum size $W$.

    $(2)$ To prove the second result, consider the scenario where $\Delta(0)=0$ (i.e., $d(0)=D-\Delta(0)=D$) and $T=D+\delta/2$ (where $\delta\to 0^+$). 
    For some $\eta>1$, and $j\in \{1,2,3,...\}$, consider the instances of packet generation process $\sigma_j=\{(\delta,W),(D(1-1/\eta^j),W/\eta^j)\}$, where a tuple $(t_i,w_i)$ denote the generation time $t_i$ and packet size $w_i$ of packet $i$.
	So, under instance $\sigma_j$, two packets are generated: $(i)$ packet $1$ of size $W$ is generated at time $\delta$, and $(ii)$ packet $2$ of size $W/\eta^j$ at time $D(1-1/\eta^j)$.
	Further, let $\sigma_0=\{(\delta,W)\}$ denote the instance of packet generation process, where a single packet is generated at time $t=\delta$ of size $W$ bits. 
	
	Note that packet $1$, i.e. $(\delta,W)$ is common in all these instances $\sigma_j$ ($j\ge 0$) of packet generation process, and if packet $1$ is delivered to the monitor until deadline $d(0)=D$, then the deadline constraint \eqref{eq:deadline-constraint} will be satisfied in the interval $[0,T]$. Therefore, if the packet generation instance is $\sigma_0$, any policy $\pi$ (causal or offline) must transmit packet $1$. 	
	However, for instances $\sigma_j$, for $j\ge1$, size of packet $2$ is smaller than the size of packet $1$, and hence, a policy may transmit packet $2$ (instead of packet $1$) for minimizing energy consumption. Note that if packet $1$ is to be delivered to the monitor before deadline $d(0)=D$, then minimum energy consumption is 
	$P(W/(D-\delta))(D-\delta)$, when packet $1$ is transmitted with constant speed $W/(D-\delta)$ over the interval $[\delta,D)$, i.e., from packet generation time $\delta$ to the deadline $d(0)=D$. Also, from Corollary \ref{cor:P(w/y)y-decreasing-y}, we know that $P(W/(D-\delta))(D-\delta)>P(W/D)D$.  On the other hand, if packet $2$ (for instance $\sigma_j$, $j\ge1$) is to be delivered to the monitor before deadline $d(0)=D$, then the minimum energy required is only $P((W/\eta^j)/(D-D(1-1/\eta^j)))(D-D(1-1/\eta^j))=P(W/D)D/\eta^j$, when packet $2$ is transmitted with constant speed $(W/\eta^j)/(D-D(1-1/\eta^j))$ from the instant the packet is generated at $D(1-1/\eta^j)$, until the deadline $d(0)=D$. Further, because $\eta>1$,  $P(W/D)D/\eta^j>P(W/D)D$ (for all $j\ge 1$). Hence, transmitting packet $2$ consumes less energy compared to packet $1$ (when packet $2$ is transmitted with appropriate speed over an appropriate interval of time).  
	
	
	However, a causal policy $\pi$ does not know the actual instance of packet generation process according to which the packets are being generated.
	So, after packet $1$ $(\delta,W)$ is generated, 
	if $\pi$ does not transmit packet $1$, waiting for packet $2$ to get generated, and if the packets are being generated according to packet generation instance $\sigma_0$, then packet $2$ will never be generated. Hence, such policy $\pi$ will not be able to satisfy the deadline constraint \eqref{eq:deadline-constraint}. So, any feasible causal policy $\pi$ must begin to transmit packet $1$ at some time $t=r_1<D$. 
	Let	
	$\pi$ begins to transmit packet $1$ at time $r_i<D$, and transmits at least $W(1-1/\eta)$ bits of packet $1$ until time $D(1-1/\eta^i)$, for some $i\ge 1$ 
	(this must be true for some $i\ge 1$, otherwise $\pi$ cannot deliver packet $1$ before $d(0)=0$ if the actual packet generation instance is $\sigma_0$; obvious by taking $i\to\infty$). 
	If the actual instance of packet generation process (according to which the packets are being generated) is $\sigma_i=\{(\delta,W),(D(1-1/\eta^i),W/\eta^i)\}$. 
	Then, $\pi$ consumes at least $P(W(1-1/\eta)/(D(1-1/\eta^i)-\delta))(D(1-1/\eta^i)-\delta)$ units of energy in interval $[\delta,D(1-1/\eta^i))$, transmitting the $W(1-1/\eta)$ bits of packet $1$ (at constant speed $W(1-1/\eta)/(D(1-1/\eta^i)-\delta)$). On the other hand, $\pi^\star$ consumes at most $P((W/\eta^i)/(D/\eta^i))(D/\eta^i)=P(W/D)(D/\eta^i)$ units of energy required to transmit packet $2$ $(D(1-1/\eta^i),W/\eta^i)$ over the interval $[D(1-1/\eta^i),D)$ with constant speed $W/D$ (as it satisfies the deadline constraint \eqref{eq:deadline-constraint}). Therefore, the competitive ratio \eqref{eq:CR-definition} of any causal policy $\pi$ is
	\begin{align} \label{eq:lb-CR-varW}
		\textsc{cr}_{\pi}&\ge\frac{P\left(\frac{W(1-1/\eta)}{D(1-1/\eta^i)-\delta}\right)(D(1-1/\eta^i)-\delta)}{P(W/D)D/\eta^i}, \nonumber \\
		&\stackrel{(a)}{>}\frac{P\left(\frac{W(1-1/\eta)}{D(1-1/\eta^i)}\right)D(1-1/\eta^i)}{P(W/D)D/\eta^i}, \nonumber \\
		&\stackrel{(b)}{\ge}\frac{P(W/D)D(1-1/\eta^i)}{P(W/D)D/\eta^i}, \nonumber \\
		&=\eta^i-1,
	\end{align}
    where we got $(a)$ using Corollary \ref{cor:P(w/y)y-decreasing-y}, and the fact that $D(1-1/\eta^i)-\delta<D(1-1/\eta^i)$ (because $\delta>0$). Also, $(b)$ holds because $P(\cdot)$ in an increasing function, and for $\eta>1$ and $i\ge 1$, $W(1-1/\eta)/(D(1-1/\eta^i))>1$. 
    
    Since the lower bound \eqref{eq:lb-CR-varW} on the competitive ratio of any causal policy $\pi$ increases with $\eta$, taking $\eta\to\infty$, we get the second result. 

\end{document}